\documentclass[11pt,a4paper]{article}
\usepackage{jheppub}
\usepackage{amsthm}
\usepackage{mathtools}
\usepackage{physics}
\usepackage{array}
\usepackage{enumitem}
\usepackage{bm}
\usepackage{hyperref}
\usepackage[capitalize]{cleveref}
\usepackage{graphicx}
\usepackage{xcolor}
\usepackage[vcentermath]{youngtab}
\newcommand{\Q}{\mathbb{Q}}
\newcommand{\C}{\mathbb{C}}
\newcommand{\R}{\mathbb{R}}

\arxivnumber{2508.11583}

\title{Anomaly cancellation for a $U(1)$ factor}
\author[a]{Ben Gripaios}
\author[b]{and Khoi {Le Nguyen Nguyen}}

\affiliation[a]{Cavendish Laboratory, University of Cambridge, \\ J.J. Thompson Avenue, Cambridge, CB3 0US, U.K.}

\affiliation[b]{DAMTP, University of Cambridge, \\ Wilberforce Road, Cambridge, CB3 0WA, U.K.}

\emailAdd{gripaios@hep.phy.cam.ac.uk}
\emailAdd{kl518@cam.ac.uk}

\abstract{We use methods of arithmetic geometry to find solutions to the abelian local anomaly cancellation equations for a four-dimensional gauge theory whose Lie algebra has a single $\mathfrak{u}_1$ summand, assuming that a non-trivial solution exists. The resulting polynomial equations in the integer $\mathfrak{u}_1$ charges define a projective cubic hypersurface over the field of rational numbers. Generically, such a hypersurface is (by a theorem of Koll{\'a}r) unirational, making it possible to find a finitely-many-to-one parameterization of infinitely many solutions using secant and tangent constructions. As an example, for the Standard Model Lie algebra with its three generations of quarks and leptons (or even with just a single generation and two $\mathfrak{su}_3\oplus\mathfrak{su}_2$ singlet right-handed neutrinos), it follows that there are infinitely many anomaly-free possibilities for the $\mathfrak{u}_1$ hypercharges. We also discuss whether it is possible to find all solutions in this way.}

\begin{document}
	\maketitle
	\flushbottom
	\section{Introduction \label{sec:intro}}
	For a gauge quantum field theory to make sense, anomalies must cancel. For even the most basic kind of anomaly, namely the local (a.k.a. perturbative) anomaly arising from chiral fermions, the problem of finding the anomaly-free representations of the gauge Lie algebra is a tough one, since it boils down to solving polynomial equations in the integers\footnote{For the abelian summand of the Lie algebra, these integers are the charges; for the semisimple summand, they are the Dynkin labels.} that label the representations.
	
	There has recently been significant progress in attacking this problem in four spacetime dimensions, beginning in \cite{Costa_2019} with the observation that in the simplest case of the Lie algebra $\mathfrak{u}_1$ (as in, {\em e.g.}, quantum electrodynamics) with an arbitrary fixed number of charges, there is a way to produce a new solution from a pair of old ones. In \cite{Allanach_2020}, it was shown that this construction is an ancient one in arithmetic geometry (where it goes by the name of either the method of chords or of secants) and that it can be used to find all solutions. In brief, the polynomial equations define a projective variety over the field $\mathbb{Q}$ of rational numbers,\footnote{Since $\mathbb{Q}$ is not algebraically complete, we should work in the setting of {\em schemes}. But we shall use the {\em lingua franca} of varieties in the hope of remaining readable by physicists.}  with the points corresponding to nonchiral (a.k.a. vectorlike) solutions forming linear subspaces; by choosing a disjoint pair of such subspaces whose dimensions sum to that of the variety itself, one can be sure that some secant between them goes through any other point in projective space, {\em ergo} every point on the variety corresponding to a chiral solution.
	
	To use the jargon,\footnote{The jargon is admittedly confusing for the initiate, owing to the fact that we have both rational numbers and rational functions in mathematics. By way of clarification, here a {\em variety over the rationals}, on the one hand, means a variety whose defining equations have coefficients lying in $\Q$, while a {\em rational point} is one whose coordinates lie in $\Q$, rather than some field extension thereof. The notions of {\em rational variety} and {\em unirational variety} that follow, on the other hand, involve rational functions and make sense for a variety over any field.}
	this argument shows that the variety is a rational variety.
	Roughly (we shall be more precise later), this means that we can use rational functions to define a bijective map from almost all of projective space\footnote{Namely, away from the vanishing loci of the denominators of the rational functions.} to almost all of the variety; this map provides a 1-1 parameterization of almost all of the points and the remaining points can be found with a little extra work.
	
	In this paper, we examine a case which is much more general: namely, we suppose that the Lie algebra corresponding to the gauge group has 
	$\mathfrak{u}_1$ as its abelian summand.\footnote{The case with an abelian summand of higher rank will be discussed elsewhere \cite{Gripaios_toappear_fano}.} In other words, it may have an arbitrary semisimple part, such as the $\mathfrak{su}_2 \oplus \mathfrak{su}_3$ of the Standard Model. This will be a sum of $n$ simple summands.
	
	Here the problem itself decouples into two parts. The first part is to ensure that the anomaly of the semisimple summand cancels. This reduces to the problem of finding the anomaly-free representations of the simple Lie algebras $\mathfrak{su}_N$, with $N \geq 3$. This part is far from being completely solved, but again there has been encouraging progress \cite{Gripaios_2024_irreducible,Gripaios_2025_products}. We will have nothing more to say about it here, supposing instead that, as our starting point, we are given an anomaly-free representation of the semisimple summand. This will be a sum of $m$ irreducible representations. 
	
	Our task, then, is to solve the second part of problem, which is to
	deal with the $\mathfrak{u}_1$ summand. Even this part is still much harder than the pure $\mathfrak{u}_1$ case, for two reasons.
	Firstly, we have $n$ additional equations in the $m$ unknown integer $\mathfrak{u}_1$ charges to satisfy, coming from the mixed anomalies between each simple summand and the $\mathfrak{u}_1$ summand. Secondly, the coefficients of the equations now depend on the choice of semisimple summand and its representation, via the integer dimensions and Dynkin indices.
	
	Nevertheless, we will see that it is possible to find infinitely many solutions for the allowed charges in cases where there exists at least one solution.\footnote{We have not found any non-trivial examples with no solutions. Such a case would certainly be interesting, but not for physics!} The reason that this is possible is that the $n$ additional equations are homogeneous and linear in the $m$ charges, so the equations may still be considered as defining a projective cubic hypersurface. These are objects of intense study in algebraic geometry and much is known. In particular, it is known that such a cubic hypersurface, while usually not a rational variety, is nevertheless typically a unirational variety (over any field!) \cite{Kollar_2002}. (The atypical cases correspond to reducible varieties or cones over elliptic curves; in both cases all rational points can be found, at least in principle.)
	
	Roughly, unirational means that we can use rational functions to define a dominant rational map from projective space to the variety; this map provides a finitely-many-to-one parameterization of infinitely many solutions.
	Unlike in the case of a rational variety, this map does not parameterize almost all of the solutions. Indeed, it is typically very far from doing so.\footnote{We thank J. Koll\'{a}r for pointing this out.} Nevertheless, one can still hope to find all solutions by repeated application of secant and tangent constructions. This is proven to work if the dimension of the hypersurface is large enough \cite{Papanikolopoulos_2017,Brandes_2021} and there is evidence that it works for the cubic surfaces we consider in examples coming from physics.\footnote{There are, however, examples of cubic surfaces where this is known not to work \cite{Siksek_2012}.}
	
	As well as allowing us to write down explicit formul\ae\ for the solutions, our methods allow us to draw an important qualitative conclusion for physics.
	Namely, we will see that a gauge theory with a single $\mathfrak{u}_1$ has infinitely many\footnote{We disregard the additional solutions trivially obtained by rescaling all charges by a constant multiple.} solutions to the anomaly cancellation equations if it has just one, provided that $m-n \geq 5$.\footnote{This is a corollary of a much stronger statement. To wit, except possibly in the special case of a cone over an elliptic curve, the closure of the rational points is a union of connected components of the real locus, in the usual euclidean topology \cite{Mazur_1992}.} To see the power of this, consider the example of the Standard Model hypercharge, where $n=2$. It is well-known \cite{Geng_1989,Weinberg_2013,Lu_2020} that with a single generation of quarks and leptons, {\em i.e.} $m=5$, there are just three possible hypercharge assignments (indeed, the problem reduces to solving a cubic equation in a single unknown). Our results immediately imply that with either more than one generation, or even just two right-handed neutrinos, there must be infinitely-many solutions, given that there is one solution, namely the observed hypercharges.\footnote{We shall see in \cref{sec:smexample} that there are also infinitely many possibilities with just a single right-handed neutrino.} 
	
	The outline is as follows. In the next section, we will describe a variety of algorithms that can be used to build an explicit parametrization of an infinite subset of points on a cubic hypersurface. As for the pure $\mathfrak{u}_1$ case, these algorithms are based on elementary geometric constructions using secants and tangents. We briefly mention an algorithm that starts from just a single point, but since any hypersurface containing a single point in fact must contain infinitely many points, it is convenient in practice to use a simpler algorithm starting from two points. We also describe even simpler algorithms starting either from a single line or a singular point (neither of which is guaranteed to exist, as the examples show). We end this Section by discussing whether one can find all solutions using such methods. In \cref{sec:su2} we explicitly describe a number of examples based on the simplest possible semisimple summand, namely $\mathfrak{su}_2$ (so $n=1$), with the smallest number of charges ($m=6$) that takes us beyond the special realm of curves to surfaces, while still allowing us to draw pretty pictures. We conclude by considering examples with the Standard Model gauge Lie algebra $\mathfrak{su}_3\oplus\mathfrak{su}_2\oplus\mathfrak{u}_1$.
	
	\section{From physics to geometry \label{sec:theory}}
	
	Denoting the possible integer charges of the aforementioned $m$ irreducible representations by $x_i$, $i \in \{1,\dots,m\}$,\footnote{We charge-conjugate right-handed fermions present in the theory so that we only have to consider a theory of left-handed fermions.} and indexing the simple summands by $j,j^\prime \in \{1,\dots,n\}$, the gauge anomaly cancellation equations are
	\begin{align}
		\sum_{i}\left(\prod_{j \neq j^\prime} T_i^{j^\prime} D^j_i \right) x_i&=0, \label{eqn:anom_lin_mixed}\\
		\sum_{i}\left(\prod_{j} D^j_i \right) x_i&=0, \label{eqn:anom_lin_grav}\\
		\sum_{i}\left(\prod_{j} D^j_i \right)x_i^3&=0, \label{eqn:anom_cubic}
	\end{align} 
	where $D^j_i$ and $T^j_i$ are respectively the dimension and Dynkin index (both of which may be taken to be integer-valued \cite{Gripaios_2025_asymp}) of the irreducible representation labelled by $i$ restricted to the simple summand $j$.\footnote{The penultimate equation is usually understood as coming from the requirement that the theory can be consistently coupled to gravity, but it can also be understood as coming from the requirement that the theory can be defined on spacetimes with non-trivial topology (see {\em e.g.} \cite{Davighi_2020}).}
	
	We have a system of $n+1$ linear equations (\cref{eqn:anom_lin_mixed} being in fact $n$ equations, one for each value of $j'$) and one cubic equation in the $m$ integers $x_i$. Since all of them are homogeneous, solving them over the ring of integers is equivalent (by clearing denominators) to solving them over the field $\Q$ and so it is convenient to work over $\Q$ in what follows, so that we can carry out the usual constructions of geometry. Homogeneity moreover implies that, given a solution $(x_1,\dots,x_m) \in \Q^m$ in which not all $x_i$ vanish, one can obtain infinitely many other solutions in a trivial way by multiplying each $x_i$ by a fixed element of $\Q$. 
	We remove this\footnote{Doing so has the additional benefit of leading to a projective variety; like the compact manifolds in differential geometry, these are by far the nicest ones.} by defining $\Bbbk P^{m-1}$, the $(m-1)$-dimensional projective space over any field $\Bbbk$, to be the set of equivalence classes $[x_1:\dots:x_m]$ under the equivalence relation
	\begin{equation}
		(x_1,\dots,x_m)\sim(y_1,\dots,y_m)\Leftrightarrow\,\exists\,k\in\Bbbk\setminus\{0\}: (x_1,\dots,x_m)=(ky_1,\dots,ky_m).
	\end{equation}
	
	As well as the case $\Bbbk = \Q$, we will later have cause to consider various extensions of $\Q$, including the real and complex numbers, denoted $\R$ and $\C$, respectively, as well as quadratic and $p$-adic extensions $\Q_p$ for a prime $p$, and the algebraic closure $\overline{\Q}$ of $\Q$.
	
	We may assume without loss of generality that the linear constraints are non-degenerate (if not, discard some). If $m<n+1$, then the system of equations is overdetermined, and generically we expect there to be only the trivial solution, while if $m=n+1$, then the linear system of equations has a unique solution which can be checked for consistency with the cubic equation. We thus focus on the case $m>n+1$. Here, we can use the linear constraints to eliminate $n+1$ variables from the cubic one to get a homogeneous cubic equation in $m-n-1$ variables which defines a \emph{cubic hypersurface} in $\Q P^{m-n-2}$, whose rational points\footnote{We shall also use the term \emph{$\Bbbk$-point} to denote points over a field $\Bbbk$, reserving the term rational point for the case $\Bbbk = \Q$.} define equivalence classes of solutions to the anomaly cancellation equations in the manner just described. 
	This cubic hypersurface is an example of a (possibly reducible, projective) \emph{variety}.\footnote{In common with most of the mathematical literature, we will insist that a variety be irreducible and use \emph{reducible variety} to cover the general case.}
	
	\subsection{Rational and unirational varieties \label{subsec:unirat}}
	
	We now define the notions of rational and unirational varieties, which play a central r\^{o}le in this work. Given two varieties $X$ and $Y$, a \emph{rational map} $\phi:X\dashrightarrow Y$\footnote{The dashed arrow indicates that the map is not necessarily defined everywhere on $X$.} is an equivalence class of morphisms defined on non-empty open\footnote{In this work, we use the Zariski topology unless we specify otherwise.} ({\em ergo} dense) sets, where two morphisms are considered equivalent if they agree on the intersection of their domains of definition.
	A rational map is \emph{dominant} if it has a representative whose image is dense in its codomain; this allows us to define composition of dominant rational maps, and we say such a map is \emph{birational} if it has a dominant rational inverse map.
	
	Now we come to the crucial definitions. We say a variety $X$ over $\Bbbk$ is \emph{rational} if there exists an $l$ and a birational map $\Bbbk P^l \dashrightarrow X$; similarly, we say a variety is \emph{unirational} if there exists an $l^\prime$ and a dominant rational map $\Bbbk P^{l^\prime} \dashrightarrow X$.\footnote{In this work, when we just say that a variety is (uni)rational without specifying the field $\Bbbk$, then we mean that it is (uni)rational over $\Q$.} In the former case $l$ must equal the dimension of the variety and in the latter case we can take $l^\prime$ to equal the dimension of the variety \cite{Kollar_2002}, so that the inverse image of a point is generically a finite set.
	
	Over an infinite field such as $\Q$, such a map (provided we can find one) allows us to parameterize infinitely many points on the variety. This parametrization is finitely-many-to-one in the case of a unirational variety and one-to-one in the case of a rational variety.
	
	Unfortunately, over a non-algebraically closed field such as $\Q$, the fact that the map is dominant does not guarantee that we are able to find all solutions in this way. Indeed, over any field $\Bbbk$ the map induced on the $\Bbbk$-points can fail to be surjective because the image of a dominant map is only required to be dense in its codomain.  But further problems can arise when $\Bbbk$ is not algebraically closed, as the following example \cite{Kollar_2004} shows. The rational map  $\Bbbk P^1\dashrightarrow\Bbbk P^1$ (which is in fact a morphism) given by $[x_1:x_2]\mapsto [x_1^2:x_2^2]$ is dominant, but while it is surjective on the $\Bbbk$-points  when $\Bbbk=\C$, it is far from being so when $\Bbbk=\R$, and the situation is even worse when $\Bbbk=\Q$. This problem is typical for unirational varieties over non-algebraically closed fields $\Bbbk$, and we will see it arise in our examples based on physics, though it does not arise if the map exhibits our variety as a rational variety. No unirational parametrization is known that gives almost all, or even a positive percentage according to some suitable counting, of the rational points on a generic cubic surface \cite{Kollar_private}, so in general we are still likely to miss most solutions. We will address this issue in \cref{subsec:mordell}.

	\subsection{Parameterizing rational points on cubic hypersurfaces \label{subsec:kollar}}
	The cubic hypersurface defined by \cref{eqn:anom_cubic,eqn:anom_lin_grav,eqn:anom_lin_mixed} can take many forms, depending on the specific values of the coefficients that are determined by the semisimple summand. An important point to be made is that we do not
	always obtain a rational nor even a unirational variety in every case. So let us discuss the possibilities that may arise and how we may find rational points in each case. 
	
	A first possibility is that the variety may be reducible. In the language of \cref{eqn:anom_cubic,eqn:anom_lin_grav,eqn:anom_lin_mixed}, this means that the cubic polynomial on the left hand side of \cref{eqn:anom_cubic} factorises when restricted to the linear suspace defined by \cref{eqn:anom_lin_grav,eqn:anom_lin_mixed}. We will see an explicit example in $\mathfrak{su}_2\oplus\mathfrak{u}_1$ gauge theory in \cref{sec:su2}.
	
	Such cases are easily dealt with, because a reducible cubic hypersurface necessarily reduces to either a union of a hyperplane and an irreducible quadric hypersurface, or to a union of three hyperplanes. In either case, we have that the irreducible components are rational varieties if and only if they have a rational point. In the case of a quadric, an explicit birational map can be obtained by constructing rational lines\footnote{In this paper, we say that a line is rational or real if the associated equation defines a line over $\Q$ and $\R$ respectively.} out of the given point and finding their other point of intersection with the quadric (which involves solving a quadratic equation in one unknown with one, {\em ergo} two, roots in $\Q$).
	
	So the first step in analysing any particular gauge theory is to determine whether the corresponding cubic hypersurface is reducible or not (this can be done using computer algebra packages such as \texttt{Mathematica} or \texttt{Macaulay2}).
	
\begin{figure}
		\centering
		\includegraphics[width=0.8\linewidth]{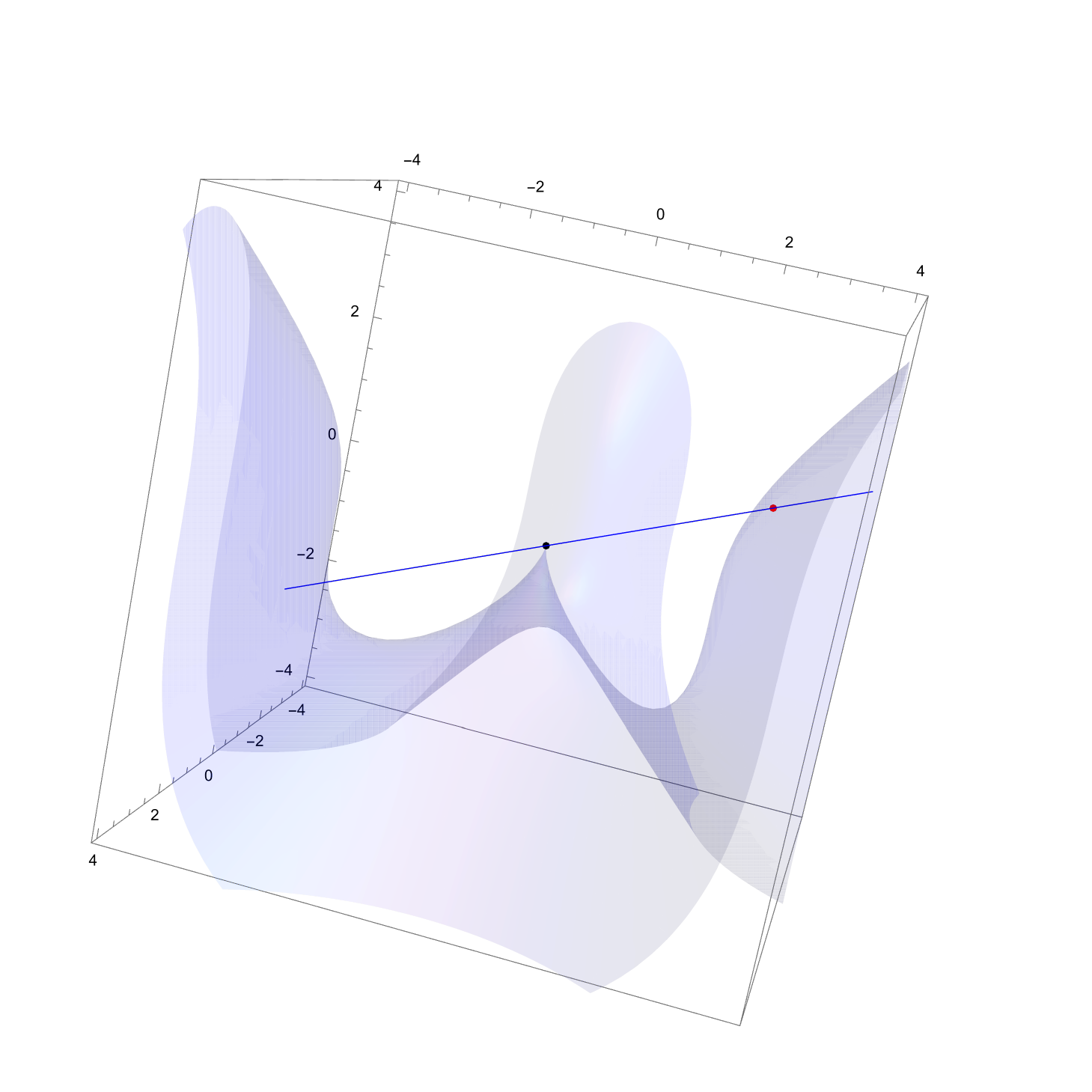}
		\caption{An affine view of a projective cubic surface which contains a rational double point, shown in black. Every rational line through this point, such as the blue one, intersects the cubic surface at only one other rational point, shown in red, unless it wholly lies on the surface. By considering all such lines, we obtain a birational map between the cubic surface and the projective plane, proving that the former is a rational variety.}
		\label{fig:example_singular_point}
	\end{figure}

	Having done so, we may now assume that the cubic hypersurface is irreducible. The next step is to ascertain whether or not it has a singular point. Again, this can be carried out with the help of the aforementioned packages.
	If it has a singular point, then the singular point is either a double point or a conical singularity. If it is a double point (an example is for pure $\mathfrak{u}_1$ gauge theory with an even number of charges, as described in \cite{Allanach_2020}; another example in $\mathfrak{su}_2\oplus\mathfrak{u}_1$ gauge theory is given in \cref{sec:su2}), the variety is rational, as can be shown again by constructing rational lines out of the given point and finding their other point of intersection with the cubic (which involves solving a cubic equation in one unknown with one repeated root in $\Q$, {\em ergo} another root in $\Q$). This is illustrated in \cref{fig:example_singular_point} for a cubic surface. If it is a conical singularity, then our variety can be described in terms of a hypersurface in one lower dimension, and we restart the whole process using that variety.
	
	We are left with the (generic) case in which our variety is smooth, having no singular points. It is clear that such a variety need be neither rational not unirational in general. Indeed, consider the simplest nontrivial case in which our hypersurface is a cubic curve. Either this curve may not have a rational point (in which case it is certainly not unirational) or it is an elliptic curve. An elliptic curve is certainly not rational, because the genus of a curve is a birational invariant and an elliptic curve has genus one while the projective line has genus zero. Moreover, it is clear that an elliptic curve need not be unirational either, because there exist elliptic curves with a finite number of rational points ({\em i.e.} those whose Mordell-Weil group is pure torsion). In fact, L{\"u}roth's theorem \cite{Luroth_1875} guarantees us that no elliptic curve can be unirational.
	
	Elliptic curves are, however, abelian varieties, and the issue of finding their rational points has been the subject of intense study. In particular, algorithms exist to find their rational points \cite{Silverman_1992}, though they are not (yet) guaranteed to terminate. Since this is a rather long story and is anyway of a very different nature as compared to our main thread, we will not say much more about it here.
	
	Thus we are left with considering a smooth irreducible cubic hypersurface that is neither an elliptic curve nor a cone thereon. Here, a remarkable theorem of Koll\'{a}r \cite{Kollar_2002} states that over any field, such a hypersurface is a unirational variety if and only if it has a rational point.\footnote{Koll{\'a}r's theorem generalizes B. Segre's analogous result for cubic surfaces over the rationals \cite{Segre_1951}.}
	
	The proof of Koll\'{a}r's theorem is via an explicit geometric construction of a suitable dominant rational map, which can therefore be used to parameterize infinitely-many anomaly-free solutions of a gauge theory.
	
	Since the construction is somewhat involved, we will begin by describing some simpler constructions that can be used. One of these constructions is in fact general over $\Q$ and uses a pair of points, but the others rely on the presence of linear subspaces in the hypersurface. The most general method among these requires just a single line, but we also describe a method using a pair of disjoint linear subspaces.
	
	All of these constructions are based on \emph{secant and tangent constructions} for cubic hypersurfaces, which rely on the simple principle that if a line is defined over the rational numbers and intersects a cubic hypersurface at two rational points, then there must be a third point of intersection that is also rational (unless if the line entirely lies on the hypersurface). If any two of these three points coincide, then the line is a tangent to the hypersurface; otherwise it is a secant.
	
	Perhaps not suprisingly, the simplicity of these methods increases as their generality decreases, so let us begin with the least general construction.
	
\begin{figure}
		\centering
		\includegraphics[width=0.8\linewidth]{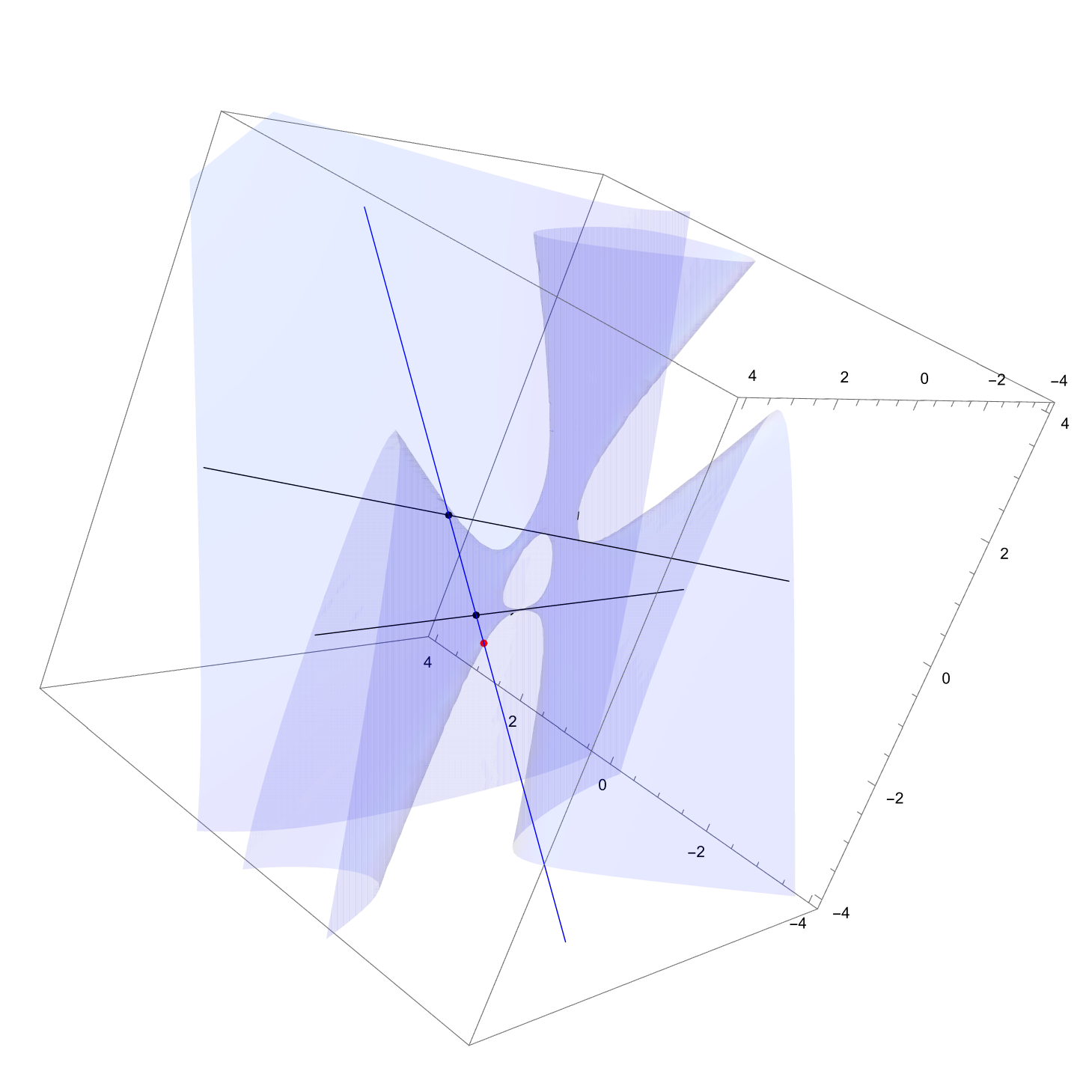}
		\caption{An affine view of a projective cubic surface containing two skew rational lines, shown in black. A secant (shown in blue) intersecting these two lines in the two rational points shown in black also intersects the cubic surface at a third rational point, shown in red, unless it wholly lies on the surface. By considering all such secants, we obtain a birational map between the cubic surface and the projective plane, proving that the former is rational.}
		\label{fig:example_two_lines}
	\end{figure}

	Suppose that our cubic hypersurface contains two disjoint linear subspaces whose dimensions sum to that of the hypersurface. In all such cases bar one, at least one such subspace must have dimension exceeding half that of the cubic hypersurface. But then the cubic hypersurface cannot be smooth \cite{Kollar_2004,Eisenbud_2016}, and so it is rational by virtue of its singular point(s), and it suffices to project from one such point. The one exception is when we have an even-dimensional cubic hypersurface that contains two disjoint linear subspaces each of half the dimension, such as a cubic surface with two skew lines. Then we can exhibit the variety as a rational one by constructing lines joining a point on one linear subspace to a point on the other and finding their third point of intersection with the cubic. This gives a birational map from a product of projective spaces to the variety, but such a product is itself birational to projective space. This method also works for the other cases considered at the start of this paragraph. The map is not defined if the secant itself lies wholly on the hypersurface, so to find all points on the cubic hypersurface we also need to consider the points lying on lines intersecting the two disjoint linear subspaces. An example is the pure $\mathfrak{u}_1$ case, as described in \cite{Allanach_2020}, and in fact the same method can be applied to (products of) irreducible representations of $\mathfrak{su}_N$ \cite{Gripaios_2024_irreducible,Gripaios_2025_products}. We illustrate this construction for a cubic surface with two skew lines in \cref{fig:example_two_lines}.
	
	It turns out that the above result still holds in certain situations where the linear subspaces are not defined over the ground field $\Bbbk$ but some extension thereof. The idea is that while a line defined over $\Bbbk$ through a single $\Bbbk$-point will not generically hit other $\Bbbk$-points in the hypersurface, if we consider a quadratic extension $\Bbbk'$ of $\Bbbk$, then such a line will hit a point in that extension and its conjugate \cite{Mordell_1969,Kollar_2004}. So assuming that we have an even-dimensional cubic hypersurface $X$ that contains a pair of disjoint linear subspaces, each of which is of half the dimension of the hypersurface, is defined over $\Bbbk'$ and is conjugate to the other over $\Bbbk$, $X$ is again a rational variety over $\Bbbk$. Indeed, such a hypersurface is certainly rational over $\Bbbk'$ by the method of secants described in the previous paragraph; if we restrict the domain to pairs of points on each linear subspace that are conjugate to each other over $\Bbbk$, then the line connecting them is defined over $\Bbbk$, and the third point of intersection is also a $\Bbbk$-point. This map is birational because through every $\Bbbk$-point on $X$, there exists a unique $\Bbbk$-line that intersects the two linear subspaces in a pair of conjugate $\Bbbk'$-points.
	
\begin{figure}
		\centering
		\includegraphics[width=0.8\linewidth]{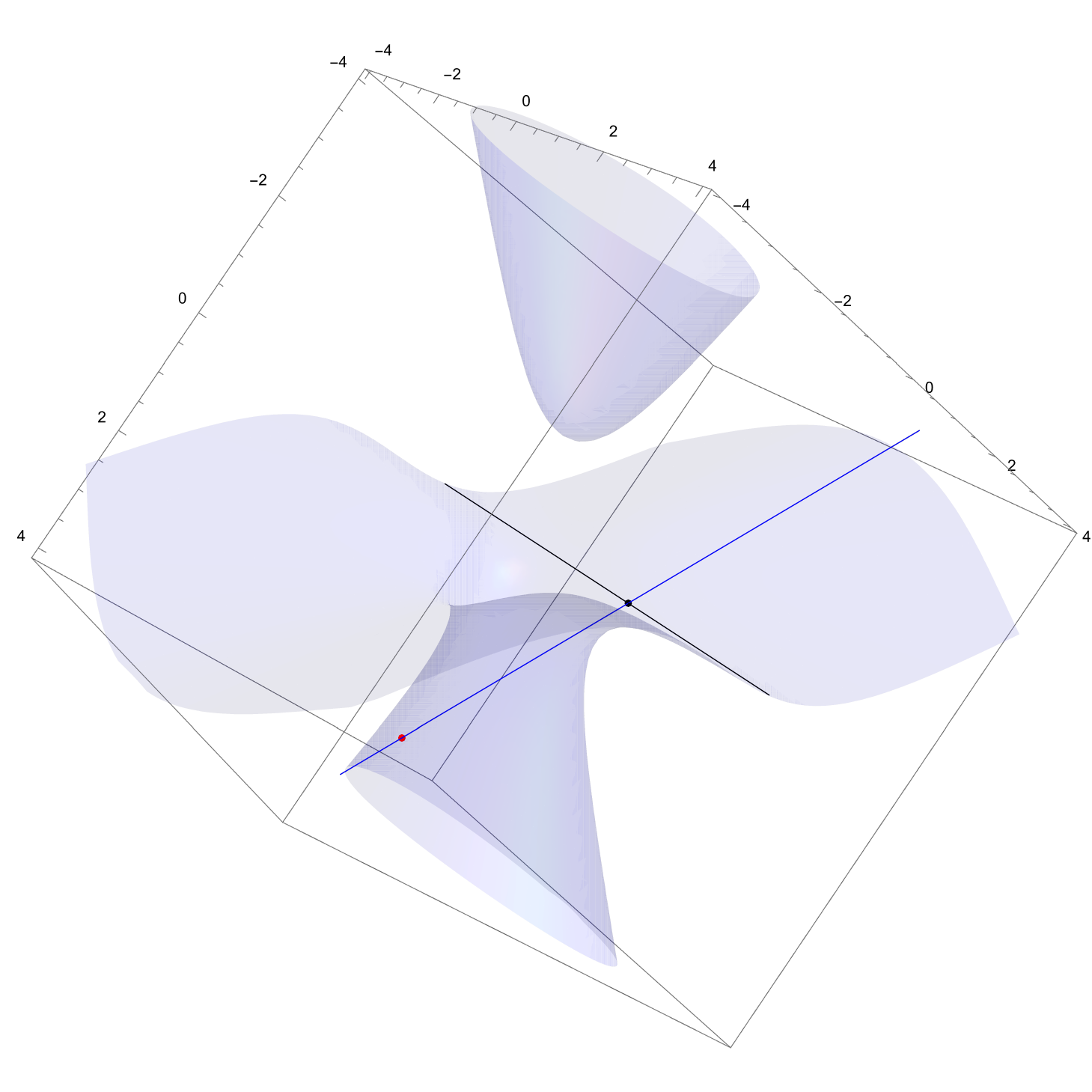}
		\caption{An affine view of a projective cubic surface containing a rational line, shown in black. The blue line is tangent to the surface at the black point (which lies on the black line) and intersects the surface at one other rational point, shown in red, unless it wholly lies on the surface. By considering all tangents along all points on the line, we obtain a rational map from the projective plane to the cubic surface, proving that such a surface is unirational.}
		\label{fig:example_one_line}
	\end{figure}

	Next, suppose that our cubic hypersurface $X$ contains a line $L$. The following construction, described in \cite{Debarre_2015} (\cite{Clemens_1972} attributes it to Max Noether), exhibits the hypersurface as a unirational variety. At each point $x$ on $L$, we consider the lines tangent to $X$. Along such a tangent line, the cubic hypersurface restricts to a cubic in one variable with a repeated rational root, so generically the line intersects $X$ in a third rational point. By considering all points along $L$ and all tangent lines at each of those points, we obtain a rational map from a product of projective spaces whose total dimension equals that of the hypersurface. To see that this map is dominant, consider the plane spanned by a point $y \in X \setminus L$ and $L$. This plane intersects $X$ in a cubic curve containing $L$, which is thus necessarily reducible and is the union of $L$ and a conic. This conic intersects $L$ in two points unless it is reducible, {\em i.e.} is a pair of lines. These two points are the preimages of $y$ under the rational map and we see that the map is two-to-one unless the point $y$ is itself on a line in $X$ that intersects $L$ (in which case it is not defined) or if the conic touches the line (in which case it is one-to-one). Again, since a product of projective spaces is birational to projective space (with the appropriate dimensions), we have shown that $X$ is unirational. An example in $\mathfrak{su}_2\oplus\mathfrak{u}_1$ gauge theory is given in \cref{sec:su2}. We illustrate this construction for a cubic surface containing a rational line in \cref{fig:example_one_line}.
	
	This explicit construction allows us to see how the problem described at the end of \cref{subsec:unirat} arises in this case: the two points of intersection of the conic and $L$ will not necessarily be defined over $\Q$, and the map defined by the construction will not be able to hit the corresponding point $y$.
	We will see this phenomenon explicitly in the examples in \cref{sec:su2}.
		
	Next, suppose we are given just a pair of rational points on $X$, which we assume are nonsingular (if either is singular, then we can use the associated construction described earlier). Given any pair of tangent lines, one through each of these points, we construct a new pair of points in $X$ by intersecting the pair of tangent lines with $X$. The line through this new pair of points intersects $X$ in a fifth point. Supposing the hypersurface to have dimension $l$, this construction gives a dominant rational map from $\Bbbk P^{l-1} \times \Bbbk P^{l-1} \dashrightarrow X$ as long as one point does not lie on the tangent space of $X$ at the other;\footnote{A proof that this rational map is dominant for a general cubic hypersurface of dimension above one can be found in \cite{Kollar_2002}; we give the proof for the case of surfaces below.} composing with a birational equivalence $\Bbbk P^{2(l-1)} \to \Bbbk P^{l-1} \times \Bbbk P^{l-1}$ exhibits $X$ as a unirational variety.
	
\begin{figure}
		\centering
		\includegraphics[width=0.8\linewidth]{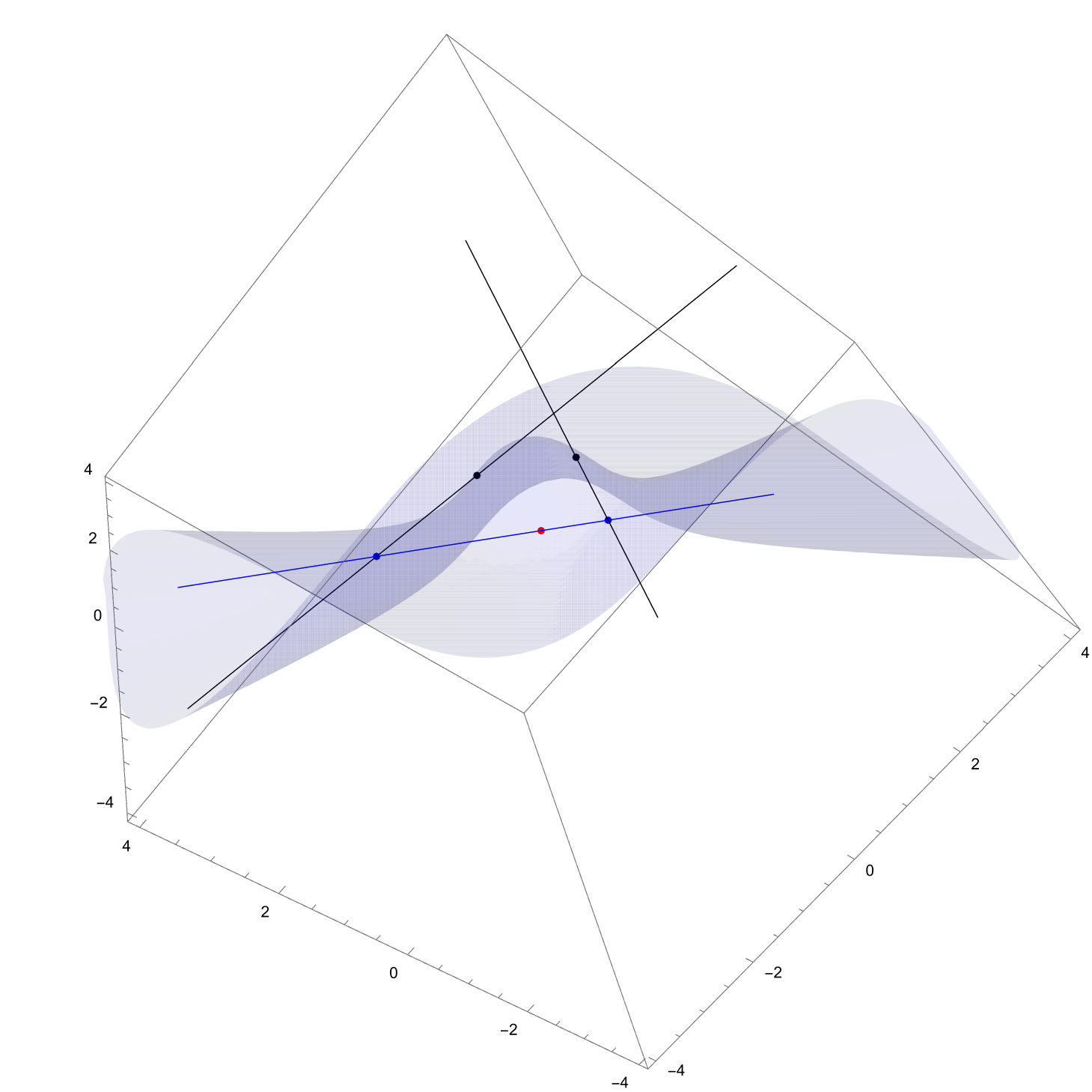}
		\caption{An affine view of a projective cubic surface which contains two rational points, shown in black. We construct two lines tangent to the surface at each of these points, which are also shown in black (they do not intersect in the Figure). Each of these tangents intersects the surface at one other rational point, shown in blue, unless it wholly lies on the surface. The line through the two blue points, also shown in blue, intersects the surface at a fifth rational point, shown in red, unless it also wholly lies on the surface. By considering all such pairs of tangent lines, we obtain a rational map from the projective plane to the cubic surface whose image is the fifth point, proving that such a surface is unirational.}
		\label{fig:example_two_points}
	\end{figure}

	Except in the case of a cubic surface ($l=2$), the parameterization obtained in this way is highly degenerate. It can be reduced straightforwardly to a finite map as described in \cite{Kollar_2002}. Since all of our examples involve cubic surfaces, we will not need to do so here. We illustrate this construction for a cubic surface containing two rational points in \cref{fig:example_two_points}.
	
	In the case where we have a cubic surface $X$ containing two rational points $p$ and $q$ but no rational line, the rational map is dominant and generically six-to-one. To see this, let us denote by $T_p$ and $T_q$ the planes tangent to $X$ at $p$ and $q$ respectively. On these two planes, the cubic surface restricts to two generically irreducible projective cubic curves $C_p$ and $C_q$, each with a double point at $p$ and $q$ respectively. These two planes intersect transversely in a line $L$ which does not lie on $X$; instead, it intersects $X$ generically at three points which, by construction, also lie on both $C_p$ and $C_q$. Given a point $r\in X$ that is not on $C_p\cup C_q$, we consider all the lines that pass through $r$ and every rational point on $C_p$. Such a line intersects $T_q$ at a third rational point; the set of all of these points is another cubic curve $C_r$ on $T_q$. But by B{\'e}zout's theorem, the coplanar cubic curves $C_q$ and $C_r$ generically intersect at nine points. Three of these are the intersections of $L$ and $X$, so we are left with six points on $C_q$ satisfying the condition that a line through $r$ and any of these six points also intersects $C_p$ at another point. These six pairs of points on $C_p$ and $C_q$ are precisely the preimages of $r$ under the rational map, which is thus generically six-to-one.\footnote{The existence of a six-to-one rational map from projective space to a smooth cubic surface containing a $\Bbbk$-point for any field $\Bbbk$ was proven by Manin in \cite{Manin_1986}. 
		Manin also showed, \emph{ibid.}, that the map is dominant if $\Bbbk$ is either an infinite field or a finite field with at least 34 elements. Koll{\'a}r's construction was shown to be such a map for an arbitrary finite field by Knecht in \cite{Knecht_2015}, and our proof for the case of $\Bbbk=\Q$ is identical.}
	
	The above arguments also show that infinitely-many rational points on a cubic hypersurface $X$ with a pair of rational points can be found using this map, together with those on lines in $X$ (if there are any) and on the intersections of $X$ with the spaces tangent to $X$ at the two given rational points. The latter are easily found because these intersections are generically irreducible cubics, each having one of the given points as a double point, so we can just project from them. Nevertheless, this parameterization is only unirational, and we still miss many rational points on $X$.
	
	In fact, all we need to exhibit unirationality is a single rational point (which may be assumed smooth). We recall that a generic $\Bbbk$-line passing through a $\Bbbk$-point on $X$ will hit a point in a quadratic extension of $\Bbbk$ and its conjugate. By applying the construction for two points just described to this pair of points, and restricting the pair of tangent lines to also be conjugate to one another, then the final point one obtains using this construction is guaranteed to be a  $\Bbbk$-point. In this way\footnote{Some technicalities must be overcome for arbitrary $\Bbbk$; we refer to \cite{Kollar_2002} for details.} one is able to obtain a dominant rational map from $\Bbbk P^{2(l-1)}$ to $X$.
	
	Koll\'{a}r's theorem implies that over $\Q$ a smooth irreducible cubic hypersurface of dimension exceeding one has infinitely many points if it has just one.\footnote{In fact, this is also true for general cubic hypersurfaces of dimensions above one, for it is true if we are reducible and if we have a cone, because every point on a cone has a line through it.} This in turn shows that our algorithm starting from just two points is in fact generally applicable over $\Q$. Since it is much simpler than Koll\'{a}r's full construction, we will use it in the examples. 
	
	Summing up, we have seen that in the reducible or rational cases we can find a one-to-one parametrization of almost all solutions to the anomaly cancellation equations \cref{eqn:anom_cubic,eqn:anom_lin_grav,eqn:anom_lin_mixed}, while in the unirational case we can find a finitely-many-to-one parameterization of infinitely many, but far from all, of them.
	
	As for the question of whether a cubic hypersurface has a rational point, the answer depends, unsurprisingly, on the dimension of the hypersurface. Specifically, Heath-Brown showed that a smooth projective cubic hypersurface of dimension at least 8 always has a rational point \cite{Heath-Brown_1983}. Before considering lower dimensions, we recall the \emph{Hasse principle}, which is the hypothesis that a projective variety has a rational point if it has an $\R$-point and a $\Q_p$-point for every prime $p$. This principle is true for quadrics (this result being the celebrated Hasse--Minkowski theorem) \cite{Cassels_1978}, a fact which was of great significance in solving the anomaly cancellation problem in two dimensions \cite{Camp_2024}. For cubic sevenfolds, Hooley showed that the Hasse principle applies \cite{Hooley_1988}. Yet, the principle is known to not hold for the simplest case of (two-dimensional) cubic surfaces.\footnote{Swinnerton-Dyer was the first to disprove the Hasse principle for cubic varieties by exhibiting two counterexamples which were smooth cubic surfaces \cite{Swinnerton-Dyer_1962}. Cassels and Guy followed up with the proof that the diagonal projective surface $5x^3+12y^3+9z^3+10t^3=0$ is yet another counterexample \cite{Cassels_1966}, and Bremner produced one such family of diagonal projective surfaces \cite{Bremner_1978}.} One obstruction to having a rational point beyond the Hasse principle is the Brauer-Manin obstruction, which is conjectured to be the only one in favourable cases \cite{Colliot-Thelene_2015}. In spite of these hurdles, very recent results \cite{Keyes_2025} show that almost all cubic hypersurfaces of dimensions between 3 and 7 inclusively contain a rational point.
	
	\subsection{Finding all solutions \label{subsec:mordell}}
	
	We have seen that for an arbitrary cubic hypersurface, it is possible to find infinitely many rational points, starting from just a single point. We now address the problem of how to find all points.
	
	In the case of a rational (or reducible) cubic hypersurface, a map which exhibits it as a rational variety already enables us to find almost all of the rational points. The missing points lie in a subvariety of lower dimension and can usually easily be found in an {\em ad hoc} way, as the examples show.
	
	The unirational case (which is the generic one) is much more challenging, because of the problem described at the end of \cref{subsec:unirat}. In particular, a map exhibiting our variety as a unirational one typically misses almost all of the rational points. The situation is not without hope, however, since given any set of rational points, such as the infinite set we already have, we can try to find further rational points by applying further secant and tangent constructions. In particular, given any two rational points, we can find a third rational point by finding the third intersection of the secant that connects them with the hypersurface, and give just one rational point we can find infinitely many more rational points by finding the other intersection of each and every tangent line to the hypersurface at that point with the hypersurface. Moreover, this process can be iterated an arbitrarily large number of times.
	
	At least in the examples, we will see some evidence that all (or at least many more) points can be obtained in this way. In particular, even performing just one iteration leads to a large multiple\footnote{As an example, for the surface defined by \cref{eqn:one_rational_line} and shown in \cref{fig:example_one_line} which contains a line, secants between 80 of the points obtained using the unirational parameterization yielded 3\,185 new points where these secants intersect the surface, an increase of a factor of order 40.} of new rational points, which when plotted at least give the appearance of being dense in the real points, as they must be.
	
	Unfortunately, not much has been proven regarding whether this process of iterating tangent and secant constructions finds all points or not, even if we iterate an arbitrarily large number of times. It was shown in \cite{Siksek_2012} that all points on a cubic surface with two disjoint rational lines\footnote{Of course, such a cubic surface is anyway a rational variety.} can be found in this way starting from just a single point and the same is true \cite{Papanikolopoulos_2017,Brandes_2021} for any cubic hypersurface whose dimension is at least 29 (which is, moreover, guaranteed to have a rational line). In the other direction, \cite{Siksek_2012} also gave examples of diagonal cubic surfaces in which a minimal starting set necessarily has infinite cardinality.
		
	Given the difficulty of finding all solutions in the unirational case, it is important to know when a cubic hypersurface is a rational variety. At least in the case of cubic surfaces, this question has been answered by Swinnerton-Dyer in \cite{Swinnerton-Dyer_1971}. Namely, a smooth cubic surface over $\Q$ is rational if and only if it has a rational point along with a set of 2, 3, or 6 skew lines that are stable under the action of the Galois group $\mathrm{Gal} \, (\overline{\Q},\Q)$. Using this, Pannekoek \cite{Pannekoek_2009} classified all smooth rational cubic surfaces into five types based on the Galois-stable orbits of the lines: a cubic surface of type (I) has two skew rational lines, type (II) surfaces have two skew lines over $\overline{\Q}$ and no set of two skew rational lines, those of types (III) and (IV) have one and two orbits of three skew lines over $\overline{\Q}$, and finally a cubic surface is of type (V) if it has a set of six skew lines over $\overline{\Q}$ that is stable under the action of the Galois group. Moreover, a rational cubic surface is of one and only one type. For surfaces of type (I) and (II), the birational parameterization of the rational points that we need is got by drawing rational secants through (conjugate) pairs of points on the two (conjugate) skew rational lines, as explained previously. For other types, no geometric construction of the map is known, but one can revert to Swinnerton-Dyer's construction (using blow-ups). In all the examples that follow where we do get a rational surface, it is always of type (I): the two skew rational lines can be either vectorlike (as encountered in past literature) or chiral (a new phenomenon, which we will explain later).
	
	Before turning to these examples, we make the following remark. We have seen that many of the constructions for finding rational points require us to understand the lines (or more generally linear subspaces) in a cubic hypersurface. It turns out that such linear subspaces play a much more significant role when it comes to physics. Indeed, as observed in the case of two spacetime dimensions in \cite{Camp_2024}, such linear subspaces are in 1-1 correspondence with anomaly-free representations of gauge Lie algebras with the same semisimple factor, but with multiple $\mathfrak{u}_1$ summands! To wit, while points correspond to solutions with a single $\mathfrak{u}_1$ summand, lines correspond to solutions with two $\mathfrak{u}_1$ summands, and so on. To study these systematically requires us to study the corresponding Fano varieties of lines, planes, {\em \&c.}, in the hypersurface. We will do so elsewhere \cite{Gripaios_toappear_fano}. 
	
	We now proceed to give concrete examples of these constructions. While our methods can be applied to cubic hypersurfaces of any dimension greater than one we have chosen to focus on the case of cubic surfaces, not least because we are able to draw pictures, but also because it shows that even in this simplest possible case a rich variety of phenomena can occur. Another motivation for focusing on the case of surfaces is that the real topology of projective cubic surfaces is completely known and classified \cite{Knorrer_1987,Holzer_2006}. The most relevant result there is that the real topology roughly correlates with the number of real lines. In particular, smooth cubic surfaces over $\R$ containing 27, 15 or 7 lines are diffeomorphic to the connected sum of 7, 5 or 3 copies of $\R P^2$ respectively, while those containing only 3 lines are diffeomorphic to either $\R P^2$ or the disjoint union of $\R P^2$ and the two-sphere $S^2$. Some of these features will be visible on the plots that follow.
	
	\section{Examples with $\mathfrak{su}_2\oplus\mathfrak{u}_1$ \label{sec:su2}}

	Consider a theory with gauge Lie algebra $\mathfrak{su}_2\oplus\mathfrak{u}_1$ where the fermions transform in the representation $\bigoplus_{i=1}^6 d_i$ of the Lie algebra $\mathfrak{su}_2$, where the positive integer $d_i$ is the dimension of the $i$-th irrep summand. Since an irrep with dimension $d_i$ has Dynkin index $d_i(d_i^2-1)/6$ (see, {\em e.g.}, \cite{Gripaios_2025_asymp}), the gauge anomaly cancellation equations for the integer charges $x_i$ for the $\mathfrak{u}_1$ summand are
	
	\begin{align}
		\sum_{i=1}^6 d_i(d_i^2-1)x_i&=0,\\
		\sum_{i=1}^6 d_ix_i&=0,\\
		\sum_{i=1}^6 d_ix_i^3&=0.
	\end{align}
	
	Representations of $\mathfrak{su}_2$ are always free from local anomalies, but not necessarily global ones: indeed, we must make sure that our choice of $d_i$ results in a theory that is free from the anomalies described in \cite{Witten_1982} and \cite{Wang_2019}.\footnote{There may be further global anomalies, depending on the choice of gauge group, as described {\em e.g.} in \cite{Davighi_2020}.} To avoid the former, there needs to be an even number of irreducible summands with dimensions 2 modulo 4, while an even number of those with dimensions 4 modulo 8 is required to avoid the latter. All of our choices of $d_1,\dots,d_6$ in the examples that follow satisfy these constraints.
	
	Let us first consider the case where $d_1=d_2=d_3=d_4=d_5=d_6$. In this case, the two linear constraints are identical, and the anomaly cancellation equations become
	\begin{equation}
		\sum_{i=1}^6 x_i=\sum_{i=1}^6 x_i^3 = 0.
	\end{equation}
	
	These equations define not a cubic surface but an irreducible cubic threefold called the \emph{Segre cubic primal}. The hypersurface has ten singular points, which can be got by permuting the coordinates of $[x_1:\dots:x_6]=[1:1:1:-1:-1:-1]$. Thus, it is a rational hypersurface, and all rational points can be found by projecting from one such singular point. This same threefold arises when we consider anomaly cancellation for a $\mathfrak{u}_1$ theory with six fermions and when we look for anomaly-free irreducible representations of $\mathfrak{su}_6$ \cite{Gripaios_2024_irreducible}. We will have much more to say about it elsewhere \cite{Gripaios_toappear_fano}.
	
	Apart from this case, we can assume without loss of generality that $d_5\neq d_6$, use the linear constraints (which will generically be non-degenerate) to express $x_5$ and $x_6$ as linear combinations of $x_{1,2,3,4}$, and substitute these expressions into the cubic constraint to get an equation defining the cubic surface in $\Q P^3\ni[x_1:x_2:x_3:x_4]$.
	
	We now proceed by considering cases of increasing generality in the choices of $d_i$, which can be equivalently thought of as partitioning them into subsets of equal values. These partitions now always give us cubic surfaces, although they may be singular and/or reducible. The first two partitions that follow, however, do give us familiar smooth irreducible cubic surfaces.
	
	\textbf{Case $5+1$}: In the case $d_1=d_2=d_3=d_4=d_5\neq d_6$, it is convenient to note that the linear constraints yield the unique solution $x_1+x_2+x_3+x_4+x_5=x_6=0$, so the cubic surface is defined by the equations
	\begin{equation}
		\sum_{i=1}^5 x_i^3 = \sum_{i=1}^5 x_i = 0. \label{eqn:clebsch}
	\end{equation}
	We recognise these equations as defining the smooth and irreducible \emph{Clebsch cubic surface} on which the 27 lines are all real \cite{Segre_1942}.\footnote{The lines on a cubic surface can be found using an algorithm in \cite{Boissiere_2007} or with the help of \texttt{Macaulay2}.} The 15 lines amongst these that are rational are all vectorlike, and we can find pairs of them which are skew (for example, the lines ${x_1+x_2=x_3+x_4=x_5=0}$ and ${x_1=x_2+x_3=x_4+x_5=0}$). Constructing all secants between points on these lines therefore yields an explicit birational map between the Clebsch cubic surface and the projective plane. This same cubic surface arises when we consider anomaly cancellation for a $\mathfrak{u}_1$ theory with five fermions and when we look for anomaly-free irreducible representations of $\mathfrak{su}_5$ \cite{Gripaios_2024_irreducible}.
	
\begin{figure}
		\centering
		\includegraphics[width=\linewidth]{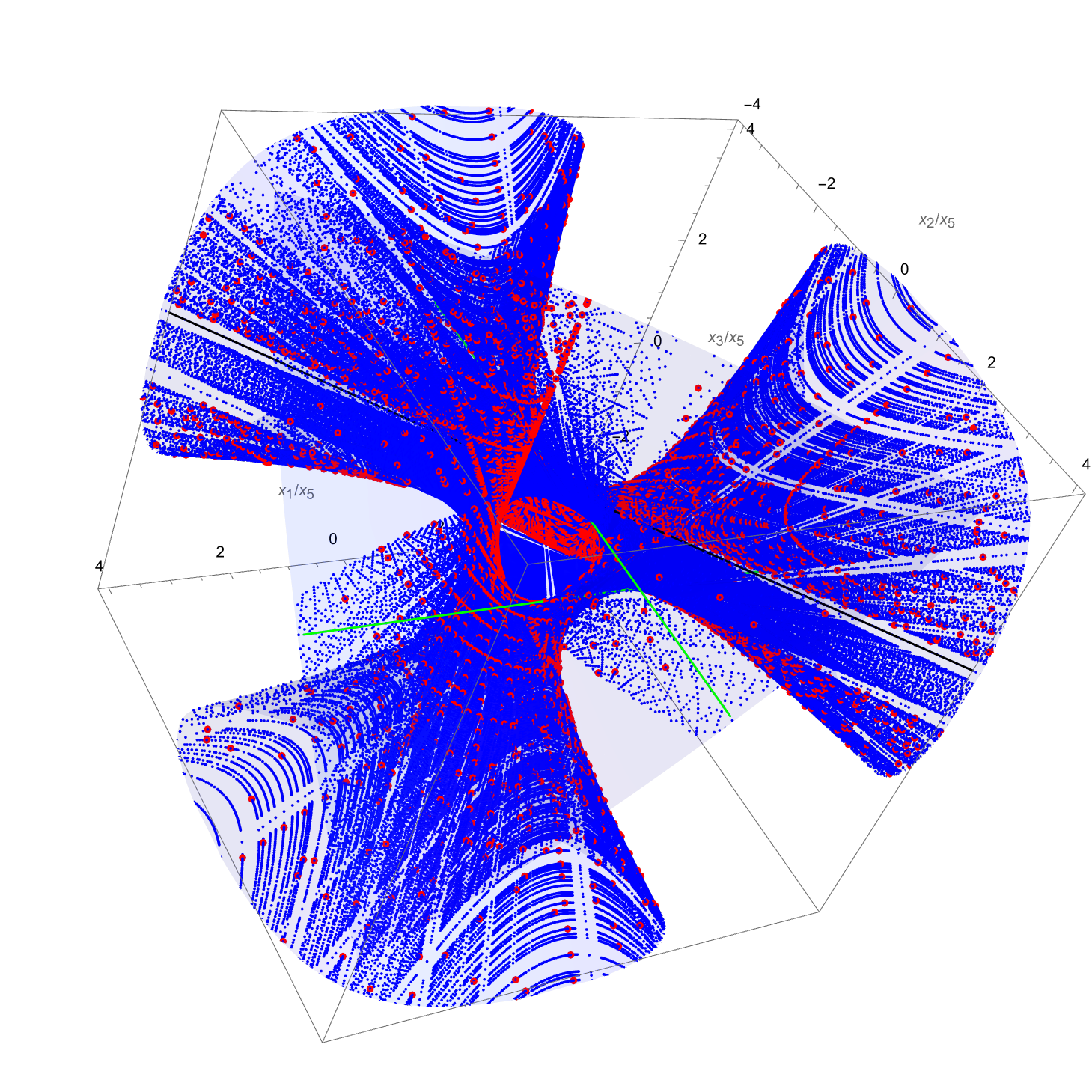}
		\caption{The Clebsch cubic surface, defined by \cref{eqn:clebsch}, as viewed in the affine patch $x_5\neq 0$. The rational points in blue are obtained by constructing secants passing through the two skew rational lines in green, and we are guaranteed to obtain all rational points on the surface in this way (together with the points on the lines on the surface). Only the rational points that are also marked in red lie on a line tangent to a point on the rational line in black, so we see that the unirational construction, while giving us infinitely many points, nevertheless misses many rational points.}
		\label{fig:debarre_clebsch}
	\end{figure}

	We use this example to illustrate how, as explained in \cref{subsec:kollar}, a unirational map hits infinitely many rational points but also misses infinitely many of them, by plotting, in \cref{fig:debarre_clebsch}, both the rational points on the Clebsch cubic surface obtained by the birational method of secants map and those points among them that could be obtained by the unirational construction starting from only one line. More details can be found in \cref{sec:clebschdetails}.
	
	\textbf{Case $3+3$}: For the partition of the six $d_i$'s into two sets of three of equal values, for instance $d_1=d_2=d_5\neq d_3=d_4=d_6$, the cubic surface is given by
	\begin{equation}
		d_1x_1x_2(x_1+x_2)+d_3x_3x_4(x_3+x_4)=0.
	\end{equation}
	This again gives us a smooth irreducible cubic surface. The surface contains 15 real lines, of which 9 are always rational and vector-like, amongst which we can find skew pairs of lines (for example, the lines $x_2=x_4=0$ and $x_1=x_3=0$). Therefore, we again obtain a rational cubic surface whose rational points can all be found using the method of secants. It is interesting, however, to consider the other 6 real lines, which take the form $[x_1:x_2:x_3:x_4]=[u:v:uc+vg:ud+vh]$ for the combinations
	\begin{itemize}
		\item $c=h=0$, $d=g=-\lambda$,
		\item $c=-d=g=\lambda$, $h=0$,
		\item $c=h=-\lambda$, $g=d=0$,
		\item $c=g=-h=\lambda$, $d=0$,
		\item $c=-d=-h=-\lambda$, $g=0$,
		\item $c=0$, $d=-g=h=\lambda$,
	\end{itemize}
	where $\lambda:=\left(\frac{d_1}{d_3}\right)^{1/3}$ and $[u:v]\in\Q P^1$. Remembering that $d_1$ and $d_3$ are distinct positive integers, we see that these 6 lines are also rational if and only if $\lambda\in\Q$, and that they do not correspond to vector-like solutions because $\lambda\neq\pm1$. We call them \emph{chiral lines}, and they are a new phenomenon which we will describe in more detail in \cite{Gripaios_toappear_fano}. In this case, we can rescale the variables $x_i$ over $\Q$ to obtain the isomorphic surface
	\begin{equation}
		y_1y_2(y_1+y_2)+y_3y_4(y_3+y_4)=0,
	\end{equation}
	for $[y_1:y_2:y_3:y_4]\in\Q P^3$. This is the \emph{equianharmonic cubic surface}, which also arises when we considered twofold anomaly-free product representations of $\mathfrak{su}_3$ \cite{Gripaios_2025_products}.
	
	The two partitions that follow, namely $4+2$ and $2+2+2$, give us reducible varieties. 
	
	\textbf{Case $4+2$}: Without loss of generality, we assume that ${d_1=d_2=d_3=d_5\neq d_4=d_6}$, for which the cubic surface is given by
	\begin{equation}
		(x_1+x_2)(x_1+x_3)(x_2+x_3)=0.
	\end{equation}
	Evidently, this variety is reducible and is the union of three non-chiral planes, each of which is obviously a rational subvariety.
	
	\textbf{Case $2+2+2$:} Let us assume that ${d_3=d_5\neq d_1=d_2\neq d_4=d_6}$ without loss of generality,  remembering that we have fixed $d_5\neq d_6$. The resulting cubic surface is
	\begin{equation}
		(x_1+x_2)[(d_3^2-d_4^2)(x_1^2-x_1x_2+x_2^2)+(d_4^2-d_1^2)(x_3^2-x_3x_5+x_5^2)+(d_1^2-d_3^2)(x_4^2-x_4x_6+x_6^2)]=0.
	\end{equation}
	where
	\begin{align}
		x_5&=-\frac{d_1(d_1^2-d_4^2)}{d_3(d_3^2-d_4^2)}(x_1+x_2)-x_3,\\
		x_6&=-\frac{d_1(d_3^2-d_1^2)}{d_4(d_3^2-d_4^2)}(x_1+x_2)-x_4.
	\end{align}
	
	\begin{sloppypar}
		Thus, the cubic surface reduces to the union of the (obviously rational) plane ${x_1+x_2=x_3+x_5=x_4+x_6=0}$ and a generically not further reducible quadric defined by the vanishing locus of the quadratic factor. The quadric contains the obvious rational points ${[x_1:x_2:x_3:x_4]=[1:-1:1:1]}$, $[-1:1:1:1]$, $[1:-1:-1:1]$ and $[1:-1:1:-1]$ (which also lie on the conic along which the plane and the quadric surfaces intersect), so it is rational, and all rational points on it can be parameterized by constructing rational lines through any of these four vectorlike rational points. Furthermore, by calculating the determinant of the quadratic form defining the quadric, we see that when $d_i=d_j+d_k$, where $\{i,j,k\}=\{1,3,4\}$, the quadric degenerates into a cone, and we can find all rational points on it simply by considering all rational lines passing through the vertex, which in this case is a rational point not lying on the plane.
	\end{sloppypar}
	
	In the remaining six partitions of 6, namely $4+1+1$, $3+2+1$, $2+2+1+1$, $3+1+1+1$, $2+1+1+1+1$ and $1+1+1+1+1+1$, the cubic surface in $\Q P^3\ni[x_1:x_2:x_3:x_4]$ that results is always generically smooth and irreducible. The first three of these partitions give cubic surfaces that contain at least one nonchiral rational line, ergo a rational point, and therefore are unirational (if it is irreducible and not a cone).
	
	\textbf{Case $4+1+1$}: Choosing without loss of generality $d_6\neq d_1=d_2=d_3=d_4\neq d_5$ (and $d_5\neq d_6$), the resulting cubic surface is given by
	\begin{equation} 
		\sum_{i=1}^4 x_i^3+\frac{d_1^2}{(d_5^2-d_6^2)^3}\left[\frac{(d_1^2-d_5^2)^3}{d_6^2}-\frac{(d_1^2-d_6^2)^3}{d_5^2}\right]\left(\sum_{i=1}^4 x_i\right)^3=0. \label{eqn:411}
	\end{equation}
	Generically, the only rational lines that are guaranteed to be contained in the surface are the vectorlike lines $x_1+x_2=x_3+x_4=0$, $x_1+x_3=x_2+x_4=0$ and $x_1+x_4=x_2+x_3=0$. These intersect pairwise, so it is not obvious at first glance that the surface is rational. Upon closer inspection, however, it may be checked by direct substitution that the surface contains the lines $x_i+x_j=x_k-(\alpha-1)x_l=0$, where $\{i,j,k,l\}=\{1,2,3,4\}$ and $\alpha$ is a root of the cubic equation
	\begin{equation}
		\alpha[(d_1^2-d_5^2)(d_1^2-d_6^2)(d_1-d_5-d_6)(d_1+d_5-d_6)(d_1-d_5+d_6)(d_1+d_5+d_6)\alpha^2-3d_5^2d_6^2(d_5^2-d_6^2)^2(\alpha-1)]=0. \label{eqn:alpha}
	\end{equation}
	
	A first observation is that the root $\alpha=0$, which is always present, gives 3 vectorlike lines which are exactly the three stated above. We therefore focus on the quadratic factor in the square brackets, whose roots $\alpha_{1,2}$ are both rational, both irrational or both complex; if the roots coincide then they must be both rational.
	
	If $d_i=d_j+d_k$ for $i,j,k\in\{1,5,6\}$, there is in fact only one root $\alpha=1$, as the quadratic expression becomes a linear one. This gives 12 rational lines which are also vectorlike; together with the other three lines, they make up the fifteen rational vectorlike lines on this cubic surface, which in this special case is in fact the Clebsch cubic surface, as can be checked by comparing \cref{eqn:clebsch,eqn:411}.
	
	If the coefficient of $\alpha^2$ in the quadratic factor is non-zero, then its roots $\alpha_{1,2}$ are controlled by the determinant, which is a quartic polynomial in $d_{1,5,6}^2$. Assuming that $\alpha_1\neq\alpha_2$, the number of rational lines on the surface is either 15 if they are rational or 3 if they are not. Moreover, in the former scenario, since neither 0 nor 1 is ever a zero of the quadratic factor, there are 12 chiral rational lines in addition to the three nonchiral ones, and we can always find a chiral line disjoint from a non-chiral one (for example, $x_1+x_2=x_3+x_4=0$ and $x_2+x_3=x_4-(\alpha_1-1)x_1=0$ are skew). Thus, we have a rational surface. For example, by setting $d_1=2$, $d_5=3$ and $d_6=7$, we get $\{\alpha_1,\alpha_2\}=\left\{7,\frac{7}{6}\right\}$. We explicitly work out the birational parameterization in this case in \cref{sec:su2exampletwolines}. We remark that determining all triples $(d_1,d_5,d_6)$ that result in the quadratic factor having two rational roots is equivalent to finding integer solutions to an inhomogeneous quartic equation in $d_1^2$, $d_5^2$ and $d_6^2$, a non-trivial task. Indeed, even the homogeneous case of this problem, which corresponds to repeated roots, is not immediately tractable.
	
	In the more general case where the two roots are not rational, we observe that while the twelve corresponding lines are definitely not rational, they are all defined over the same quadratic extension of $\Q$ (by the square root of the determinant) where $\alpha_1$ and $\alpha_2$ are conjugate elements. However, it is obvious from the form of the lines that two skew lines are not conjugate, and two conjugate lines are not skew. Not only does this mean that we cannot obtain a birational parameterization of the surface over $\Q$ using a skew pair of these lines, but it also means that there is no subset of these twelve lines that satisfies Swinnerton-Dyer's criterion for rationality. Nevertheless, we were not able to prove that once the remaining (generically complex) twelve lines are accounted for, all such surfaces are not rational.
	
	\textbf{Case $3+2+1$}: Assuming $d_6\neq d_1=d_2=d_3\neq d_4=d_5$ (and $d_5\neq d_6$), the resulting cubic surface is given by
	\begin{equation}
		d_1(x_1^3+x_2^3+x_3^3)+d_4x_4^3+\frac{d_1^3(d_1^2\!-\!d_4^2)^3(x_1\!+\!x_2\!+\!x_3)^3}{d_6^2(d_4^2-d_6^2)^3}-\frac{[d_1(d_1^2\!-\!d_6^2)(x_1\!+\!x_2\!+x_3)\!+\!d_4(d_4^2\!-\!d_6^2)x_4]^3}{d_4^2(d_4^2-d_6^2)^3}=0.
	\end{equation}
	
	This surface is only guaranteed to have the three vectorlike rational lines $x_1+x_2=x_3=0$, $x_1+x_3=x_2=0$ and $x_2+x_3=x_1=0$, which are concurrent at the Eckardt point $x_1=x_2=x_3=0$. Again, it is not obvious that the surface is rational. Nonetheless, can we still find a sufficient condition for this surface to be rational, as in the case $4+1+1$? We can. The surface in fact contains lines of the form $x_i+x_j=x_k+\beta x_4=0$, where $\{i,j,k\}=\{1,2,3\}$ and $\beta$ is a root of the cubic equation
	\begin{equation}
		\beta[(d_1^2-d_4^2)(d_1-d_4-d_6)(d_1+d_4-d_6)(d_1-d_4+d_6)(d_1+d_4+d_6)\beta^2+3d_1d_4d_6^2(d_1^2-d_6^2)\beta-3d_4^2d_6^2(d_4^2-d_6^2)]=0.
	\end{equation}
	
	Again, we observe that this equation always has the root $\beta=0$ which corresponds to the three nonchiral rational lines above, and so we focus on the quadratic factor in the square brackets. This factor becomes a linear one when $d_i=d_j+d_k$ for $i,j,k\in\{1,4,6\}$, where we have the single rational root $\beta=\frac{d_4(d_4^2-d_6^2)}{d_1(d_1^2-d_6^2)}$. Thus, we get three extra chiral rational lines. It can be further checked that in this case, there are also three new vectorlike rational lines of the form $x_i+x_j=x_4=0$, where $i,j\in\{1,2,3\}$ and $i\neq j$, for a total of nine rational lines. Of these lines, we can find two that are disjoint (for example, $x_1+x_2=x_3+x_4=0$ and $x_2+x_3=x_4=0$), and so the surface is rational in this case.
	
	If the coefficient of $\beta^2$ in the square brackets is non-zero, then the problem becomes analogous to the $4+1+1$ case again: the determinant is a quartic polynomial in $d_{1,4,6}^2$, and the roots $\beta_{1,2}$ are both rational, both irrational or both complex if they are distinct, and both rational if they coincide; moreover, $\beta_{1,2}\neq0$. Focusing on the case where there are two distinct rational roots, which gives 6 extra chiral rational lines, we can always find a pair of lines that are skew among the nine rational lines, for example $x_1+x_2=x_3=0$ and $x_2+x_3=x_1+\beta_1 x_4=0$. Therefore, our surface is rational. An example of this case is $d_1=3$, $d_4=4$ and $d_6=5$, which gives $\{\beta_1,\beta_2\}=\left\{\frac{2}{5},\frac{14}{5}\right\}$. The application of the method of secants in this case is similar to the analogous case of the $4+1+1$ partition, so we have not done it explicitly here.
	
	If the two roots are both irrational or both complex, however, then we run into the same problem as previously, namely that even though the corresponding lines are defined over the same quadratic extension of $\Q$, there are no two that are both disjoint and conjugate. In such cases, we are generically not able to proceed further than giving a unirational parameterization based on constructing lines tangent to one of the three rational lines on the surface.
	
	\textbf{Case $2+2+1+1$}: Setting $d_1=d_2$, $d_3=d_4$ and keeping $d_1$, $d_3$, $d_5$ and $d_6$ all distinct, we find a cubic surface described by the equation
	\begin{align}
		&d_1(x_1^3+x_2^3)+d_3(x_3^3+x_4^3) \nonumber \\
		&-\frac{[d_1(d_1^2\!-\!d_6^2)(x_1\!+\!x_2)\!+\!d_3(d_3^2\!-\!d_6^2)(x_3\!+\!x_4)]^3}{d_5^2(d_5^2\!-\!d_6^2)^3}+\frac{[d_1(d_1^2\!-\!d_5^2)(x_1\!+\!x_2)\!+\!d_3(d_3^2\!-\!d_5^2)(x_3\!+\!x_4)]^3}{d_6^2(d_5^2\!-\!d_6^2)^3}=0,
	\end{align}
	on which the only rational line that is guaranteed is the nonchiral $x_1+x_2=x_3+x_4=0$. Let us again seek a sufficient condition for the surface to be rational. By substitution, it can be verified that the surface contains the lines $x_3+x_4=x_i-(\alpha-1)x_j=0$, where $\{i,j\}=\{1,2\}$ and $\alpha$ is a root of \cref{eqn:alpha}, as well as the lines $x_1+x_2=x_k-(\alpha'-1)x_l=0$, where $\{k,l\}=\{3,4\}$ and $\alpha'$ is a root of \cref{eqn:alpha} but with $d_1$ replaced by $d_3$. Again, the roots $\alpha=0$ and $\alpha'=0$ return the nonchiral line above. To find more lines, we follow the exact same argument as in the $4+1+1$ case. Firstly, we get the two nonchiral rational lines $x_3+x_4=x_1=0$ and $x_3+x_4=x_2=0$ when $d_1\in\{d_5+d_6,|d_5-d_6|\}$, but these two lines intersect the line $x_1+x_2=x_3+x_4=0$ at the Eckardt point $[0:0:1:-1]$. Similarly, we get the two nonchiral rational lines $x_1+x_2=x_3=0$ and $x_1+x_2=x_4=0$ if the positive integer $d_3$ takes up one of the two values above, but again the three rational lines would be concurrent at the Eckardt point $[1:-1:0:0]$. However, if we arrange the values of the $d_i$'s such that both conditions are simultaneously satisfied, then we get all four extra lines above, amongst which we can find a pair that are skew (take $x_1+x_2=x_3=0$ and $x_3+x_4=x_1=0$, for example). The resultant surface is then rational. We can play the same game with setting the dimensions of the irreducible factors such that we get chiral rational lines instead of nonchiral ones, as was done in the $4+1+1$ case, to get more examples of rational surfaces.
	
\begin{figure}
		\centering
		\includegraphics[width=\linewidth]{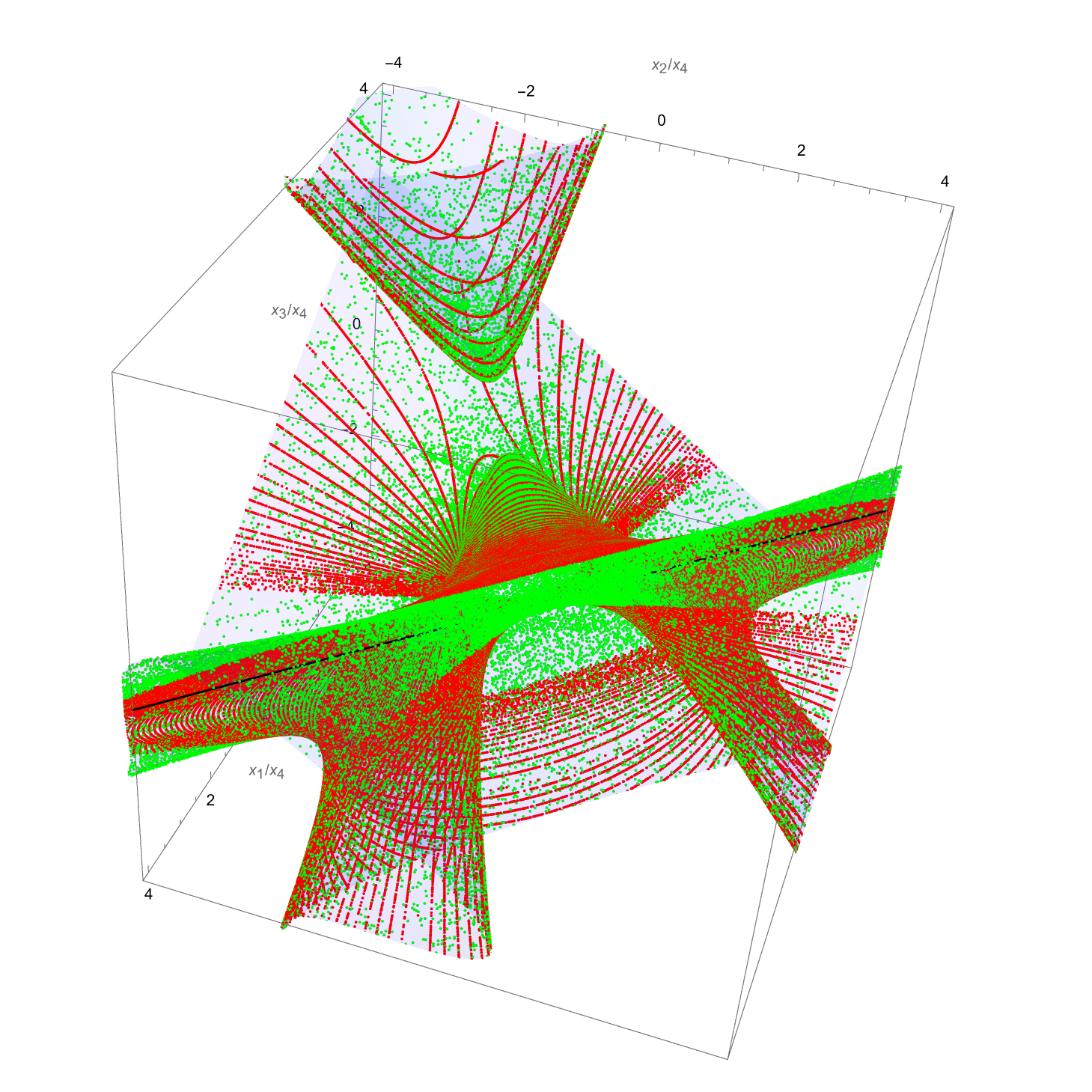}
		\caption{A subset of the rational points on the affine patch $x_4\neq0$ of the cubic surface defined by \cref{eqn:one_rational_line} which results from cancelling the gauge anomaly of the $\mathfrak{su}_2\oplus\mathfrak{u}_1$ theory where the fermions transform in the representation $2\oplus2\oplus3\oplus3\oplus7\oplus8$ (the integers being the dimensions of the irreducible summands of the representation of $\mathfrak{su}_2$). Each point in red, green or black gives a solution for the charges of the first four summands under the abelian factor, with the other two charges found from \cref{eqn:anom_lin_grav,eqn:anom_lin_mixed}. The red points were obtained by applying the unirational construction described in \cref{subsec:kollar} to the nonchiral rational line $x_1+x_2=x_3+x_4=0$, which is shown in black. This is the only rational line on this surface, which also contains 14 other real (but not rational) lines (not shown). Each green point was obtained by constructing either the secant through a pair of red points or a tangent to the surface at one of them; only a subset of the red points were used in this plot.}
		\label{fig:one_rational_line}
	\end{figure}

	Generically, however, such a surface is unirational, and it only contains the nonchiral rational line $x_1+x_2=x_3+x_4=0$. We use one such example to illustrate the parameterization based on intersecting the surface with the tangent plane at every point on the line that was previously described in \cref{sec:theory}. Specifically, in the case $(d_1,d_2,d_3,d_4,d_5,d_6)=(2,2,3,3,7,8)$, we get the smooth irreducible cubic surface
	\begin{equation}
		2(x_1^3+x_2^3)+3(x_3^3+x_4^3)-\frac{1}{49}[8(x_1+x_2)+11(x_3+x_4)]^3+\frac{1}{8}[3(x_1+x_2)+4(x_3+x_4)]^3=0, \label{eqn:one_rational_line}
	\end{equation}
	which has only one rational line of the aforementioned form. The surface also does not have a Galois-stable orbit of 2, 3 or 6 skew lines over $\overline{\Q}$, and so it is not rational. We give more details on the explicit unirational construction in \cref{subsec:kollar} in \cref{sec:su2exampleoneline}, which, together with one iteration of the algorithm in \cref{subsec:mordell}, we have used to obtain the rational points on this surface as shown in \cref{fig:one_rational_line}.
	
	Reducing the symmetry of the cubic surface further, we find surfaces that generically contain no rational line but nevertheless are still unirational by virtue of having at least one vectorlike point. Such cases are much more complicated, and we have made no attempt to deduce conditions for them to be rational. 
	
	\textbf{Case $3+1+1+1$}: If $d_1=d_2=d_3$ only and $d_1$, $d_4$, $d_5$ and $d_6$ are all distinct, the cubic surface that results always contains the three vectorlike rational poins $[1:-1:0:0]$, $[1:0:-1:0]$ and $[0:1:-1:0]$. 
	
	\textbf{Case $2+1+1+1+1$}: If we generalize further to the case where only $d_1=d_2$, then the cubic surface still contains the vectorlike rational point $[1:-1:0:0]$. 
	
\begin{figure}
		\centering
		\includegraphics[width=\linewidth]{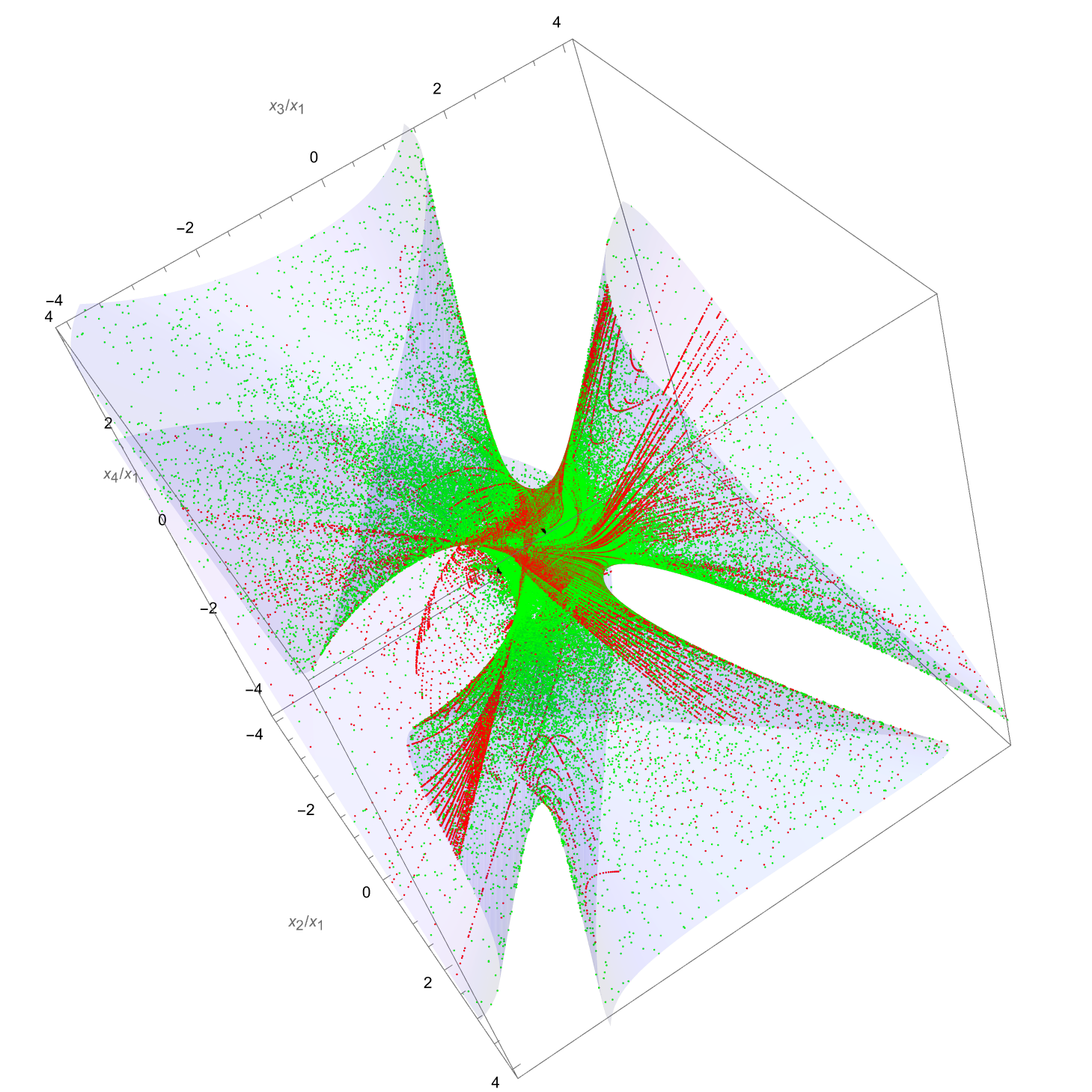}
		\caption{A subset of the rational points on the affine patch $x_1\neq0$ of the cubic surface defined by \cref{eqn:one_rational_point} which results from the cancelling the gauge anomaly of the $\mathfrak{su}_2\oplus\mathfrak{u}_1$ theory where the fermions transform in the representation $2\oplus3\oplus5\oplus6\oplus7\oplus11$ (the integers being the dimensions of the irreducible summands of the representation of $\mathfrak{su}_2$). Each point in red, green or black gives a solution for the charges of the first four summands under the abelian factor, with the other two charges found from \cref{eqn:anom_lin_grav,eqn:anom_lin_mixed}. The red points were obtained using the two rational points $[x_1:x_2:x_3:x_4]=[5:0:0:1]$ and $[4:0:-3:0]$, which are shown in black. The surface is not rational. It contains no rational line, but it does contain 27 real lines (not shown). Each green point was obtained by constructing either the secant through a pair of red points or a tangent to the surface at one of them; only a subset of the red points were used in this plot.}
		\label{fig:two_rational_points}
	\end{figure}

	\begin{sloppypar}
		\textbf{Case $1+1+1+1+1+1$}: Finally, we look at the most general case where all the $d_i$'s are distinct. Sometimes, we may find examples of the resulting cubic surface that contain a rational line: the choice $(d_1,d_2,d_3,d_4,d_5,d_6)=(2,3,5,6,7,9)$ yields a cubic surface containing only three rational lines that are all chiral and intersect pairwise, for example.
		However, the existence of lines is not guaranteed, and in general we do not expect there to be any rational line. Nevertheless, we have not found any example where there fails to be a rational point either. Thus, we conclude by presenting the case with ${(d_1,d_2,d_3,d_4,d_5,d_6)=(2,3,5,6,7,11)}$, which yields the cubic surface
		\begin{equation}
			2x_1^3+3x_2^3+5x_3^3+6x_4^3-\frac{(39x_1+56x_2+80x_3+85x_4)^3}{84\,672}+\frac{(15x_1+20x_2+20x_3+13x_4)^3}{209\,088}=0.\label{eqn:one_rational_point}
		\end{equation}
		It contains no rational line, and its 27 complex lines form a single orbit under the action of the Galois group $\mathrm{Gal}(\overline{\Q},\Q)$, so it is not rational. However, the surface does contain the two rational points $[5:0:0:1]$ and $[4:0:-3:0]$, neither of which lies on the plane tangent to the surface at the other point. Thus, we can use them to produce a dominant rational parameterization of infinitely many rational points on the surface, which can then be used to obtain more rational points by secant and tangent constructions, and we have plotted some of the rational points obtained in this manner in \cref{fig:two_rational_points}. Since the construction involved in this case is similar to that detailed in \cref{sec:su2exampleoneline} but results in much more complicated expressions, we have not explicitly written it down here.
	\end{sloppypar}
	
	\section{Standard Model-like examples \label{sec:sm}}
	We now consider anomaly cancellation in theories with gauge algebra $\mathfrak{su}_3\oplus\mathfrak{su}_2\oplus\mathfrak{u}_1$, as in the Standard Model. In particular, we consider theories with fermions of charges $x_1,\dots,x_{M_1}$ transforming in $M_1$ copies of the defining representations of both simple summands, fermions of charges $x_{M_1+1},\dots,x_{M_1+M_2}$ transforming in $M_2$ copies of the singlet of $\mathfrak{su}_3$ and the doublet of $\mathfrak{su}_2$, fermions of charges $x_{M_1+M_2+1},\dots,x_{M_1+M_2+M_3}$ transforming in $M_3$ copies of the $\mathfrak{su}_3$ antitriplet and $\mathfrak{su}_2$ singlet, and fermions of charges $x_{M_1+M_2+M_3+1},\dots,x_{M_1+M_2+M_3+M_4}$ transforming in $M_4$ copies of the singlets of both simple summands.\footnote{We remind the reader that any right-handed fermion is charge conjugated to a left-handed one.} To make sure that the $\mathfrak{su}_3$ summand is free of gauge anomalies, we require $2M_1=M_3$. These theories are of particular interest to us because the Standard Model itself is one such example, having $M_1=M_2=M_4=1$ and $M_3=2$ for a single generation. In fact, certain families of these theories that yield rational cubic hypersurfaces have been investigated in \cite{Bhattacharya_2022}, but now we have the power tools needed to consider any generic case, so long as the anomaly cancellation equations obtained from \cref{eqn:anom_cubic,eqn:anom_lin_grav,eqn:anom_lin_mixed} have a rational point.
	
	\begin{sloppypar}
		Let us first consider some scenarios where the hypersurface is zero- or one-dimensional. With just one Standard Model generation, the cubic hypersurface consists of the points $[x_1:\dots:x_5]={[0:0:1:-1:0]}$, ${[1:-3:2:-4:6]}$ and ${[1:-3:-4:2:6]}$, the latter two corresponding to the Standard Model itself \cite{Geng_1989,Weinberg_2013}. With an extra right-handed neutrino of charge $-x_6$ in the $\mathfrak{su}_3\oplus\mathfrak{su}_2$ singlet, the cubic hypersurface that results is reducible as the product of three lines in $\Q P^2$, giving us the solutions $[x_1:\dots:x_6]={[0:0:k:-k:l:-l]}$, ${[k:-3k:l:-2k-l:2k-l:4k+l]}$ and ${[k:-3k:l:-2k-l:4k+l:2k-l]}$ for ${[k:l]\in\Q P^1}$. If we add another fermion transforming in the singlet-doublet representation, then we get an elliptic curve containing only nine rational points \cite{Lu_2020}.
	\end{sloppypar}
	
	We end by turning our attention to the cases where the hypersurface is of dimension at least two. For example, we get a cubic surface by introducing two right-handed neutrinos, a fivefold if instead the Standard Model has two generations of fermionic matter, and a tenfold if there are three. We know that all of these cubic hypersurfaces contain one rational point which is got by setting the charges of the matter in one generation to their values in the Standard Model and the charge of any extra matter to zero, so we always have a unirational variety by Koll{\'a}r's theorem. These examples are considered in more detail in \cref{sec:smexample}.
	
	\acknowledgments{We thank J{\'a}nos Koll{\'a}r for feedback on an earlier draft. This work was partially supported by STFC consolidated grant ST/X000664/1 and a Trinity-Henry Barlow Scholarship.}
	
	\appendix
	
	\section{A cubic surface with two skew rational lines, one chiral and the other vectorlike \label{sec:su2exampletwolines}}
	The anomaly cancellation \cref{eqn:anom_cubic,eqn:anom_lin_grav,eqn:anom_lin_mixed} for the $\mathfrak{su}_2\oplus\mathfrak{u}_1$ theory where the fermions transform in the representation $2\oplus 2\oplus 2\oplus 2\oplus 3\oplus 7$ (the integers being the dimensions of the irreducible summands of the representation of $\mathfrak{su}_2$) give rise to the projective cubic surface
	\begin{equation}
		X_A:F_A=49(x_1^3+x_2^3+x_3^3+x_4^3)-31(x_1+x_2+x_3+x_4)^3=0 \label{eqn:xa}
	\end{equation}
	in $\Q P^3\ni[x_1:x_2:x_3:x_4]$, where $x_{1,2,3,4}$ are the charges of the fermions in the first four irreducible factors. The surface contains, among others, two rational skew lines: the vectorlike $L:x_1+x_2=x_3+x_4=0$ and the chiral $L':x_2+x_3=x_4-6x_1=0$. Given a point $x=[k:-k:l:-l]\in L$ and another point $x'=[k':l':-l':6k']\in L'$, the secant connecting them is given by $\alpha x+\beta x'$ for $[\alpha:\beta]\in\Q P^1$, and it intersects the cubic surface at points satisfying
	\begin{equation}
		\alpha\beta\left\{\alpha[k^2(k'+l')+l^2(6k'-l')]+\beta[k'^2(k-36l)-l'^2(k-l)]\right\}=0.
	\end{equation}
	
	The point with $\beta=0$, respectively $\alpha=0$, is $x$, respectively $x'$. Thus, the third point of intersection is given by the unique value of $[\alpha:\beta]$ that causes the linear factor in the braces to vanish, except for the following two cases:
	\begin{enumerate}
		\item If either the coefficient of $\alpha$ or of $\beta$, but not both, is zero, then the third point has $\beta=0$, respectively $\alpha=0$, and thus is $x$ or $x'$. The secant is tangent to the surface at the corresponding point on $L$, respectively $L'$.
		\item If both the coefficients of $\alpha$ and of $\beta$ vanish, then the secant in fact wholly lies on the surface. This occurs for the following five lines: the three rational lines $x_1+x_4=x_2+x_3=0$, $x_1+x_2=x_4-6x_3=0$ and $x_2-6x_1=x_3+x_4=0$, and the pair $2(16\pm3\sqrt{149})x_1+(67\pm\sqrt{149})x_2+50x_3=(67\pm\sqrt{149})x_1+2(16\pm3\sqrt{149})x_2+50x_4=0$, which are conjugate and irrational.
	\end{enumerate}
	
	As an example, if $k_1=k_2=1$, $l_1=-1$ and $l_2=-2$, then $[\alpha:\beta]=[29:-7]$ and we get the solution
	\begin{equation}
		[x_1:x_2:x_3:x_4:x_5:x_6]=[88:-60:-172:-52:147:-7],
	\end{equation}
	which can be checked to satisfy \cref{eqn:anom_cubic,eqn:anom_lin_grav,eqn:anom_lin_mixed} (equivalently, \cref{eqn:xa}).
	\section{A cubic surface containing only one rational line \label{sec:su2exampleoneline}}
	The anomaly cancellation \cref{eqn:anom_cubic,eqn:anom_lin_grav,eqn:anom_lin_mixed} for the $\mathfrak{su}_2\oplus\mathfrak{u}_1$ theory where the fermions transform in the representation $2\oplus 2\oplus 3\oplus 3 \oplus 7 \oplus 8$ (the integers being the dimensions of the irreducible summands of the representation of $\mathfrak{su}_2$) give rise to the projective cubic surface
	\begin{equation}
		X_B:F_B=2(x_1^3+x_2^3)+3(x_3^3+x_4^3)-\frac{1}{49}[8(x_1+x_2)+11(x_3+x_4)]^3+\frac{1}{8}[3(x_1+x_2)+4(x_3+x_4)]^3=0 \label{eqn:xb}
	\end{equation}
	in $\Q P^3\ni[x_1:x_2:x_3:x_4]$, where $x_{1,2,3,4}$ are the charges of the fermions in the first four irreducible factors. The only rational line contained in the surface is the non-chiral line $L:x_1+x_2=x_3+x_4=0$. Given a point $x=[k:-k:l:-l]\in L$, the tangent space to $X$ at $x$ is given by
	\begin{equation}
		\pdv{F_B}{x_1}\Biggr|_x (x_1-k)+\pdv{F_B}{x_2}\Biggr|_x (x_2+k)+\pdv{F_B}{x_3}\Biggr|_x (x_3-l)+\pdv{F_B}{x_4}\Biggr|_x (x_4+l)=0,
	\end{equation}
	which reduces, for $l\neq0$, to 
	\begin{equation}
		x_3+x_4=-\frac{2k^2}{3l^2}(x_1+x_2).
	\end{equation}
	
	Hence, a line $L'$ tangent to $X$ at $x$ is given by $[k+a b_1:-k+a b_2:l+a b_3:-l+a b_4]$, with 
	\begin{equation}
		b_3+b_4=-\frac{2k^2}{3l^2}(b_1+b_2). \label{eqn:b4}
	\end{equation}  
	
	We find that along $L'$, \cref{eqn:xb} restricts to a cubic in the rational variable $\frac{a}{l}$ with a double root at $\frac{a}{l}=0$ (corresponding to the point $x$) and a third root given by a rational function in $b_1$, $b_2$, $b_3$ and $\frac{k}{l}$. We also check that $l=0$ (\emph{i.e.} when this parameterization is not well-defined) corresponds to two conjugate real lines wholly contained in $X_B$ which intersect at $[1:-1:0:0]$. This parameterization was used to obtain the points in red in \cref{fig:one_rational_line}.
	
	As an example, if $b_1=b_2=b_3=\frac{k}{l}=1$, then $b_4=\frac{7}{3}$, $\frac{a}{l}=-\frac{392}{305}$ and we get the solution
	\begin{equation}
		[x_1:x_2:x_3:x_4:x_5:x_6]=[261:2091:261:-1829:-224:196],
	\end{equation}
	which can be checked to satisfy \cref{eqn:anom_cubic,eqn:anom_lin_grav,eqn:anom_lin_mixed} (equivalently, \cref{eqn:xb}).

	\section{A unirational parameterization does not reach all rational points on a cubic surface \label{sec:clebschdetails}}
	The anomaly cancellations \cref{eqn:anom_cubic,eqn:anom_lin_grav,eqn:anom_lin_mixed} for the $\mathfrak{su}_2\oplus\mathfrak{u}_1$ theory where the fermions transform in a representation $d_1\oplus d_1\oplus d_1\oplus d_1\oplus d_1\oplus d_6$, where $d_1$ and $d_6$ are distinct positive integers that denote the dimensions of the irreducible summands of the representation of $\mathfrak{su}_2$, give rise to the Clebsch cubic surface
	\begin{equation}
		X_C: F_C=x_1^3+x_2^3+x_3^3+x_4^3-(x_1+x_2+x_3+x_4)^3=0 \label{eqn:xc}
	\end{equation}
	in $\Q P^3\ni[x_1:x_2:x_3:x_4]$, where $x_{1,2,3,4}$ are the charges of the fermions in the first four irreducible factors. All 27 lines on the surface are real, and 15 of them are rational. The rational lines are all nonchiral and take the form $x_i+x_j=x_k+x_l=x_m=0$, where $\{i,j,k,l,m\}=\{1,2,3,4,5\}$ and $x_5$ is the charge of the fermion in the fifth irreducible factor. Taking for example the two skew lines $L:x_1=x_2+x_3=x_4+x_5=0$ and $L':x_1+x_4=x_2+x_5=x_3=0$, we construct secants between them to obtain the birational parameterization of a rational point on the Clebsch cubic surface as $\alpha x+\beta x'$, where $x=[0:k:l:-k]\in L_1$, $x'=[k':l':0:-k']\in L_2$ and $[\alpha:\beta]=[-ll'^2+k(l'^2-k'^2):k^2(k'-l')+l^2l']\in\Q P^1$ as was done in \cref{sec:su2exampletwolines}, missing only points on five lines on the surface that intersect both $L$ and $L'$. These two lines are coloured green, and the points parameterized thus coloured blue, in \cref{fig:debarre_clebsch}.
	
	Let us take another line on the surface, say $L'':x_1+x_2=x_3+x_4=x_5=0$, and construct the unirational parameterization by intersecting the tangent plane at every point $[k:-k:l:-l]$ on the line with the surface as was done in \cref{sec:su2exampleoneline}. If we do so, then we find that we can only hit points $[x_1:x_2:x_3:x_4]$ on the surface that satisfy
	\begin{equation}
		k^2(x_1+x_2)+l^2(x_3+x_4)=0.
	\end{equation}
	Since $k$ and $l$ are both rational, we obviously miss very many points on the surface. In \cref{fig:debarre_clebsch}, the line $L''$ is coloured black, and the red points are those obtained by constructing secants through points on $L$ and $L'$ that can also be reached by tangents at points on $L''$. As described in the main text, in this case all points can be reached by repeated construction of tangents and secants starting from these points.

	\section{Cubic hypersurfaces from Standard Model-like examples \label{sec:smexample}}
	The nomenclature in this section follows that in \cref{sec:sm}.
	
	If we have just one generation of matter in the Standard Model but two extra right-handed neutrinos of charges $-x_6$ and $-x_7$ transforming in the $\mathfrak{su}_3\oplus\mathfrak{su}_2$-singlet, then we get a cubic surface in $\Q P^3\ni[x_1:x_3:x_5:x_6]$ defined by
	\begin{equation}
		6x_1(2x_1-x_3)(4x_1+x_3)-(x_5+x_6)(36x_1^2-6x_1(x_5+x_6)+x_5x_6)=0, \label{eqn:sm_twoneutrinos}
	\end{equation}
	where $x_2=-3x_1$, $x_4=-2x_1-x_3$ and $x_7=6x_1-x_5-x_6$. The cubic surface is smooth and irreducible with 15 rational lines, of which we can find two that are disjoint (for example $x_1=x_5+x_6=0$ and $2x_1-x_3-x_5=x_6=0$), so it is rational, and rational points can be parameterized by constructing secants intersecting two such lines, in the same way as was done in \cref{sec:su2exampletwolines}. Doing so, we see that the required birational parameterization is
	\begin{equation}
		[x_1:x_3:x_5:x_6]=\alpha[0:k:l:-l]+\beta[k':l':2k'-l':0],
	\end{equation}
	where $[k:l],[k',l']\in\Q P^1$ and 
	\begin{equation}
		[\alpha:\beta]=[12kk'(k'+l')-l(2k'-l')^2:l^2(2k'-l')-6k^2k']\in\Q P^1.
	\end{equation}
	The parameterization fails when the secant is actually one of the following five lines on the surface that intersect the above two rational lines: the three rational lines $x_1=x_6=0$, $2x_1-x_3-x_5=4x_1+x_3-x_6=0$ and $x_3-2x_1=x_5+x_6=0$, and the conjugate pair of irrational lines $2(5\pm\sqrt{5})x_1+(1\mp\sqrt{5})x_3-2x_5=4(2\pm\sqrt{5})x_1-(1\mp\sqrt{5})x_3-2x_6=0$.
	
	As an example, if $k=k'=1$, $l=2$ and $l'=-1$, then $[\alpha:\beta]=[-3:1]$ and we get the solution
	\begin{equation}
		[x_1:x_2:x_3:x_4:x_5:x_6:x_7]=[1:-3:-4:2:-3:6:3],
	\end{equation}
	which can be checked to satisfy \cref{eqn:sm_twoneutrinos}. Setting $l=0$, $l'=-4k'$ and $kk'\neq0$ gives $[\alpha:\beta]=[6k:k']$ and the solution
	\begin{equation}
		[x_1:x_2:x_3:x_4:x_5:x_6:x_7]=[1:-3:2:-4:6:0:0],
	\end{equation}
	corresponding to the Standard Model.
	
	With more than one generation of matter, we know that the cubic hypersurface that results is always unirational. It turns out that they also contain linear subspaces of dimension one less than the number of generations. To obtain these subspaces, we set the ratio of the charges in each generation to that of the single-generation Standard Model, but then allow the ratios of the tuples of charges between each generation to be independent. This means that the simpler unirationality construction based on the existence of a rational line can be used to parameterize infinitely many of the rational points on the cubic hypersurface. From this subset of rational points, we can get infinitely many others by repeated application of secant and tangent procedures.
	\bibliography{unirational_references_submission_v2} 
\end{document}